\documentclass[11pt]{article}

\usepackage{amsmath,amssymb,amsthm,cancel,hyperref,graphicx,xcolor}
\usepackage{mathrsfs,wasysym}
\usepackage{booktabs}
\usepackage{tikz}
\usetikzlibrary{calc}
\usepackage{placeins}

\usepackage{fullpage}
\usepackage{slashed}
\usepackage{graphicx}

\usepackage{soul}

\DeclareRobustCommand{\Sec}[1]{Sec.~\ref{#1}}
\DeclareRobustCommand{\Secs}[2]{Secs.~\ref{#1} and \ref{#2}}

\DeclareRobustCommand{\Eq}[1]{Eq.~(\ref{#1})}

\DeclareRobustCommand{\InRef}[1]{Ref.~\cite{#1}}
\DeclareRobustCommand{\Refs}[1]{Refs.~\cite{#1}}

\newcommand{\be}{\begin{eqnarray}}
\newcommand{\ee}{\end{eqnarray}}
 
\allowdisplaybreaks

\usepackage{color}
\definecolor{darkblue}{rgb}{0,0,0.5}
\definecolor{darkgreen}{rgb}{0,0.5,0}

\title{Flavor Factorization at Two-Loops}
\author{Andrew J.~Larkoski}
\date{%
    {\it Department of Physics and Astronomy, University of California, Los Angeles, CA 90095, USA}\\
{\it Mani L. Bhaumik Institute for Theoretical Physics, University of California, Los Angeles, CA 90095, USA}\\
{\it Email:} \url{larkoa@gmail.com}\\[2ex]
    \today
}

\begin{document}
\maketitle

\begin{abstract}
\noindent Recently, a factorization theorem was proposed for partonic flavor evolution as defined by the net flavor of the Winner-Take-All axis of a jet.  We validate the factorization theorem through explicit calculation at two-loop order, and in the process extract all anomalous dimensions and renormalization factors for any ultraviolet-to-infrared flavor transition at this order.  These results can then be used to extract the renormalized hard function for flavored jet production at next-to-next-to-leading order for any process of interest.
\end{abstract}

\clearpage

\tableofcontents

\section{Introduction}

The partonic flavor of a jet's initiating particle is strongly correlated with the partons of the initial state, or the flavor of a resonance, or the species of particles from which the jet recoils.  It is therefore desirable, for both Standard Model and Beyond the Standard Model applications, to have a theoretically well-defined definition of jet flavor in which predictions can be made and tested for rates of given jet flavors.  A naive definition of jet flavor, as simply the net sum of flavors of all constituents of jets, loses infrared safety starting at next-to-next-to-leading order, and therefore cannot be used in state-of-the-art calculations.  A first definition of infrared and collinear safe jet flavor was proposed many years ago \cite{Banfi:2006hf}, as a modification of the $k_T$ clustering algorithm \cite{Catani:1993hr,Ellis:1993tq} to be aware of the flavor of the particles clustered at each step.  Recently, several other jet flavor definitions have been proposed \cite{Caletti:2022hnc,Caletti:2022glq,Czakon:2022wam,Gauld:2022lem,Caola:2023wpj} that can be applied to jets defined with the anti-$k_T$ algorithm \cite{Cacciari:2008gp}, as is ubiquitously used by modern experiments.

Of the recent jet flavor proposals, most are fully infrared and collinear (IRC) safe (at least through next-to-next-to-leading order) which enables straightforward interface with numerical integration and subtraction schemes.  However, IRC safety does not ensure a simple calculation and in some cases requires reclustering the entire event with a flavor-aware jet algorithm.  The proposal of \InRef{Caletti:2022glq}, however, maintains infrared safety but forgoes collinear safety, defining the flavor of a jet to be the flavor that flows exactly along the direction of its Winner-Take-All (WTA) recombination scheme axis \cite{Bertolini:2013iqa,Larkoski:2014uqa,gsalamwta}.  With this definition of jet flavor, the evolution of flavor from the ultraviolet to the infrared is described by a fragmentation process, similar to DGLAP evolution \cite{Dokshitzer:1977sg,Gribov:1972ri,Altarelli:1977zs}.  Recently, Ref.~\cite{Larkoski:2023upz} proposed a factorization theorem for this WTA flavor definition, and validated it through one-loop order and in particular flavor channels at two-loops.

Here, we extend the analysis of \InRef{Larkoski:2023upz} through two-loop order, calculating all anomalous dimensions and thereby validating the factorization theorem through this order.  To do this, we calculate bare jet functions which requires both tree-level $1\to 3$ as well as one-loop $1\to 2$ splitting functions and then determine their renormalized versions by absorbing divergences into multiplicative renormalization factors.  At every stage of calculations, there are numerous checks on the factorization theorem; for example, all soft divergences explicitly cancel in the bare jet functions ensuring that evolution is indeed strictly collinear in nature.  Further, in some particular flavor channels, soft quarks can contaminate hard gluon flavor, but such a term is always associated with a scaleless rapidity-divergent integral, and therefore with rapidity regularization \cite{Chiu:2011qc,Chiu:2012ir} again explicitly vanishes.  The corresponding renormalization factors can then be used to renormalize the process-dependent hard function for WTA flavor, producing predictions for flavor evolution through next-to-next-to-leading order with next-to-leading-logarithmic resummation for any desired process, interfacing with recent high precision flavored calculations \cite{Czakon:2020coa,Gauld:2020deh,Catani:2020kkl,Hartanto:2022ypo,Gauld:2023zlv}.

This paper is organized as follows.  In \Sec{sec:review}, we review the WTA flavor algorithm and the flavor fragmentation factorization theorem.  In \Sec{sec:summary}, we summarize the central results of this paper; namely, anomalous dimensions through two-loop order and the associated renormalization factors.  In \Sec{sec:example}, we present a detailed worked example of calculation of the anomalous dimension $\gamma_{g\to q}$, which illustrates the general and necessary techniques for evaluation of any flavor transition.  Intermediate results, but with many fewer calculational steps presented, for bare jet functions of all other flavor transition processes are presented in \Sec{sec:barejet}.  We conclude in \Sec{sec:concs} and appendices collect the one-loop results necessary for renormalization and extraction of two-loop anomalous dimensions.

\section{Review of Flavor Fragmentation Factorization}\label{sec:review}

In this section, we review the WTA flavor factorization theorem, starting with a review of the flavor algorithm.  This review will be brief, and we refer to the original papers \cite{Caletti:2022glq,Larkoski:2023upz} for more details.

Given a jet or any set of particles of interest, the WTA flavor algorithm proceeds as follows:
\begin{enumerate}
\item Recluster the jet with the $k_T$ algorithm \cite{Catani:1993hr,Ellis:1993tq} with the Winner-Take-All recombination scheme \cite{Bertolini:2013iqa,Larkoski:2014uqa,gsalamwta}.  At every stage of clustering, the mother particle lies in the direction of its harder/more energetic daughter, with an energy that is the sum of its daughters.\footnote{This is always true for massless daughters; for massive particles the difference between energy and magnitude of three-momentum can lead to pathologies, so typically implementations of the WTA recombination scheme compare magnitudes of three-momentum \cite{Cacciari:2011ma}.}  For jets at hadron colliders, transverse momentum with respect to the beam is compared and recombined, rather than energy.
\item Continue clustering until all particles are clustered into a single trunk of the branching history.  The direction of the resulting meta-particle necessarily lies along the direction of an individual particle in the jet.  The flavor of this particle is then defined to be the WTA flavor label of the jet.
\end{enumerate}
In a scale-invariant theory, particle production continues to arbitrarily small scales, so the algorithm above assumes that there is some cutoff $k_\perp^2$ below which particle production ceases.  In the realistic case of the QCD parton shower, that scale is comparable to the scale of hadron masses, $k_\perp^2 \sim 1\text{ GeV}^2$, and all particles at the end of the perturbative parton shower are isolated within a transverse momentum region of size $~k_\perp^2$.  In the factorization theorem, which only describes perturbative flavor evolution, this scale $k_\perp^2$ must be explicit, and the WTA flavor is defined as the sum of all parton flavors that lie within $k_\perp^2$ of the WTA axis of the jet.  For a flavor sum of multiple partons, gluons have neutral or 0 flavor, while a quark and its anti-quark have opposite flavor, but two quarks $q,q'$ such that $q' \neq \bar q$ have incompatible flavors and their sum is simply denoted as $(qq')$.

From this definition of WTA flavor, the cross section $\sigma_{pp\to f}$ for single-inclusive jet production of WTA flavor $f$ in proton-proton collision events factorizes into the form
\begin{align}
\sigma_{pp\to f} = \sum_{j} H_{pp\to j}(\{s,p_T,R,\dotsc\},\mu^2)\,J_{j\to f}(\mu^2,k_\perp^2)\,.
\end{align}
Here, $H_{pp\to j}$ is the process-dependent hard function that describes jet production of WTA flavor $j$ at the renormalization scale $\mu^2$ and is in principle sensitive to the collision energy $s$, the jet's energy or transverse momentum $p_T$, the jet radius $R$, etc., and any other parameters used to define the process.  The jet function $J_{j\to f}$, by contrast, is universal and describes collinear evolution of the flavor from $j$ at scale $\mu^2$ to the measured flavor $f$ at the physical scale $k_\perp^2$.  Assuming that the high scales are sufficiently greater than any partonic mass, the intermediate flavor sum is over all flavors $j$ considered as massless partons.

The cross section $\sigma_{pp\to f}$ is physical and as such is independent of the renormalization scale $\mu^2$, which then implies strong constraints on the hard and jet functions individually.  Collinear divergences are renormalized by matrix renormalization factors $Z_{j\to k}$ describing flavor evolution from $j$ in the ultraviolet to $k$ in the infrared as:
\begin{align}
&H_{pp\to k}^{(\text{bare})} = \sum_j H_{pp\to j}^{(\text{ren})} Z^{-1}_{j\to k}\,, &J_{k\to f}^{(\text{bare})} = \sum_j Z_{k\to j}J_{j\to f}^{(\text{ren})} \,.
\end{align}
Here, we have explicitly denoted bare and renormalized jet and hard functions and suppressed arguments for brevity.  Note that the renormalization factors for the hard and jet functions, $Z_{j\to k}$, are inverses as matrices with indices $j,k$, and act on the renormalized functions either on the right (in the infrared) or on the left (in the ultraviolet).  However, unless otherwise denoted, hard and jet functions are always assumed to be their renormalized versions.  Consistency of the factorization theorem requires that the renormalization factors multiply as matrices into the identity:
\begin{align}
\sum_j Z^{-1}_{i\to j}Z_{j\to k} = \delta_{ik}\,.
\end{align}

Demanding that the cross section is independent of $\mu^2$ leads to the renormalization group equations for the hard and jet functions, where
\begin{align}
&\mu^2\frac{d}{d\mu^2} H_{pp\to k} = -\sum_j H_{pp\to j}\gamma_{j\to k}\,, &\mu^2\frac{d}{d\mu^2} J_{k\to f} = \sum_j \gamma_{k\to j} J_{j\to f}\,.
\end{align}
Here, $\gamma_{j\to k}$ is the anomalous dimension for the flavor transition $j\to k$.  Then, given anomalous dimensions and boundary values for the hard and jet functions, predictions for the cross section for flavored jet production can be produced straightforwardly.

\section{Summary of Main Results}\label{sec:summary}

We now collect the main results of this paper for ease of reference.  In \Secs{sec:example}{sec:barejet}, the two-loop bare jet functions are calculated, and then, using the renormalization prescription, the renormalized jet function can be evaluated and its anomalous dimensions extracted.  From that procedure, we find the anomalous dimensions through two-loop order are
\begin{align}
\gamma_{g\to q}& = \frac{\alpha_s}{2\pi}\frac{T_R}{3}+\left(\frac{\alpha_s}{2\pi}\right)^2T_R\left[
C_F\left(
-\frac{377}{288}-\frac{25}{36}\log 2+\frac{\pi^2}{9}-1.276
\right)\right.\\
&\hspace{4cm}\left.+\,C_A\left(
\frac{11}{54}-\frac{11}{9}\log 2+\frac{\pi^2}{18}+0.738
\right)
\right]+{\cal O}(\alpha_s^3)\,,\nonumber\\
\gamma_{q\to g}& = \frac{\alpha_s}{2\pi}C_F\left(
-\frac{5}{8} + 2\log 2
\right)\\
&\hspace{1cm}+\left(\frac{\alpha_s}{2\pi}\right)^2C_F\left[
C_F\left(
\frac{21}{64}+\frac{45}{32}\log 2+\frac{13}{16}\pi^2+\frac{\log^2 2}{2}+\frac{\pi^2}{3}\log 2+\frac{8}{3}\log^3 2-\frac{9}{2}\zeta_3+1.17
\right)\nonumber\right.\\
&\hspace{2cm}\left.+\,C_A\left(
-\frac{65}{72}+\frac{431}{24}\log 2-\frac{4}{9}\pi^2-\frac{31}{6}\log^2 2-\frac{5}{3}\pi^2\log 2+4\zeta_3 - 1.499
\right)
\right]+{\cal O}(\alpha_s^3)\,,\nonumber\\
\gamma_{q\to q'}& = \left(\frac{\alpha_s}{2\pi}\right)^2C_F T_R\left(
\frac{43}{144}-\frac{107}{36}\log 2+\frac{\pi^2}{9}+\frac{2}{3}\log^2 2-0.0446
\right)+{\cal O}(\alpha_s^3)\,,\\
\gamma_{q\to \bar q}& =\left(\frac{\alpha_s}{2\pi}\right)^2C_F\left[
T_R\left(
\frac{43}{144}-\frac{107}{36}\log 2+\frac{\pi^2}{9}+\frac{2}{3}\log^2 2-0.0761
\right)-0.00758\left(C_F-\frac{C_A}{2}\right)
\right] \\
&\hspace{1cm}+{\cal O}(\alpha_s^3)\,.\nonumber
\end{align}
There are two other anomalous dimensions for diagonal flavor evolution, $\gamma_{q\to q}$ and $\gamma_{g\to g}$.  Because probability is conserved along the evolution, there are two sum rules of the anomalous dimensions:
\begin{align}
&0=\gamma_{q\to q} +\gamma_{q\to g}+2(n_f-1)\gamma_{q\to q'}+\gamma_{q\to \bar q}\,, &0=\gamma_{g\to g} +2n_f \gamma_{g\to q}\,.
\end{align}
Additionally, all other anomalous dimensions corresponding to other partonic flavor transitions are related by the symmetries of perturbative QCD (such as charge conjugation).  The one-loop anomalous dimensions were previously presented in \Refs{Caletti:2022glq,Larkoski:2023upz}, but the two-loop results are new.  The numerical factors represent the result from numerical integration and whose uncertainty is of the order of the final displayed digit. Two-loop anomalous dimensions are sufficient for collinear evolution at next-to-leading logarithmic accuracy.

To extract these anomalous dimensions, we needed the renormalization factors.  Through two-loops, these are
\begin{align}
Z_{g\to q} (\mu_0^2,\mu^2)&= \frac{\alpha_s}{2\pi} \left(
\frac{\mu_0^2}{\mu^2}
\right)^\epsilon\frac{T_R}{\epsilon}
\frac{1}{3}\\
&\hspace{-1cm}+\left(
\frac{\alpha_s}{2\pi}
\right)^2\left(
\frac{\mu_0^2}{\mu^2}
\right)^{2\epsilon}\frac{T_R}{\epsilon^2}\left[
C_F\left(
\frac{5}{48}-\frac{\log 2}{3}+\epsilon\left(
-\frac{377}{576}-\frac{25}{72}\log 2+\frac{\pi^2}{18}-0.6378
\right)
\right)\right.\nonumber\\
&\hspace{3cm}\left. +\, C_A\left(
\frac{11}{36}+\epsilon\left(
\frac{11}{108}-\frac{11}{18}\log 2+\frac{\pi^2}{36}+0.369
\right)
\right)-\frac{2}{9}n_f T_R
\right]+{\cal O}(\alpha_s^3)\,,\nonumber\\
Z_{q\to g}(\mu_0^2,\mu^2)&=\frac{\alpha_s}{2\pi}\left(
\frac{\mu_0^2}{\mu^2}
\right)^\epsilon\frac{C_F}{\epsilon}\left(
-\frac{5}{8}+2\log 2\right)+\left(
\frac{\alpha_s}{2\pi}
\right)^2\left(
\frac{\mu_0^2}{\mu^2}
\right)^{2\epsilon}\frac{C_F}{\epsilon^2}\\
&\hspace{-2cm}\times\left[
C_F\left(
-\frac{1}{2}\left(
\frac{5}{8}-2\log 2
\right)^2+\,\epsilon\left(
\frac{21}{128}+\frac{45}{64}\log 2+\frac{13}{32}\pi^2+\frac{\log^2 2}{4}+\frac{\pi^2}{6}\log 2+\frac{4}{3}\log^3 3-\frac{9}{4}\zeta_3+0.585
\right)
\right)\right.\nonumber\\
&\hspace{-1cm}+C_A\left(
\frac{11}{12}\left(
-\frac{5}{8}+2\log 2
\right)+\epsilon\left(
-\frac{65}{144}+\frac{431}{48}\log 2-\frac{2}{9}\pi^2-\frac{31}{12}\log^2 2-\frac{5}{6}\pi^2\log 2+2\zeta_3-0.749
\right)
\right)\nonumber\\
&\hspace{-1cm}\left.+\,n_f T_R\left(\frac{5}{12}-\frac{4}{3}\log 2\right)
\right]+{\cal O}(\alpha_s^3)\,,\nonumber\\
Z_{q\to q'}(\mu_0^2,\mu^2) &= \left(
\frac{\alpha_s}{2\pi}
\right)^2\left(
\frac{\mu_0^2}{\mu^2}
\right)^{2\epsilon}\frac{C_F T_R}{\epsilon^2}\left[
-\frac{5}{48}+\frac{\log 2}{3}+\epsilon\left(
\frac{43}{288}-\frac{107}{72}\log 2+\frac{\pi^2}{18}+\frac{\log^2 2}{3}-0.0223
\right)
\right]\nonumber\\
&\hspace{1cm}+{\cal O}(\alpha_s^3)\,,\\
Z_{q\to \bar q}(\mu_0^2,\mu^2) &= \left(
\frac{\alpha_s}{2\pi}
\right)^2\left(
\frac{\mu_0^2}{\mu^2}
\right)^{2\epsilon}\frac{C_FT_R}{\epsilon^2}\left[
-\frac{5}{48}+\frac{\log 2}{3}\right.\\
&\hspace{1cm}\left.+\,\epsilon\left(
\frac{43}{288}-\frac{107}{72}\log 2+\frac{\pi^2}{18}+\frac{\log^2 2}{3}-0.0381-0.00379\frac{C_F - \frac{C_A}{2}}{T_R}
\right)
\right]+{\cal O}(\alpha_s^3)\,.\nonumber
\end{align}
Again, they satisfy sum rules for probability conservation and now we have
\begin{align}
&1=Z_{q\to q}+Z_{q\to g}+2(n_f-1)Z_{q\to q'}+Z_{q\to \bar q}\,,&1 = Z_{g\to g}+2n_f Z_{g\to q}\,,
\end{align}
These renormalization factors can then be used to renormalize bare hard functions from a calculation of WTA flavored jet production through next-to-next-to-leading order.

\section{Worked Example: $\gamma_{g\to q}$ at Two-Loops}\label{sec:example}

In this section, we present the details of calculation of the bare jet function for the flavor transition $g\to q$, $J_{g\to q}^{\text{(bare)}}(\mu_0^2,k_\perp^2)$, at two-loop order.  This example illustrates all of the interesting phenomena that arise in the calculation of WTA flavor jet functions in general and so can be used to interpret the results we present in the following section.

The bare jet function will have contributions from both tree-level $1\to 3$ splitting as well as one-loop $1\to 2$ splitting.  The master formula for the bare jet function from $1\to 3$ splitting is given by the expression
\begin{align}
J^{(2,\text{bare})}_{j\to f}(\mu_0^2,k_\perp^2) = \int d\Phi_3\, |{\cal M}|^2\,\Theta_\text{WTA}\,,
\end{align}
for the flavor transition $j\to f$, where $d\Phi_3$ is differential three-body phase space, $|{\cal M}|^2$ is the matrix element for the splitting, and $\Theta_\text{WTA}$ is the WTA measurement function.  In what follows, we will work with dimensionless phase space and matrix elements, explicitly dividing or multiplying by appropriate factors of the jet energy.  Differential collinear three-body phase space in $d=4-2\epsilon$ dimensions we work with can be expressed as \cite{Gehrmann-DeRidder:1997fom,Ritzmann:2014mka}
\begin{align}
d\Phi_3 = \frac{4^{1-\epsilon}}{(4\pi)^{5-2\epsilon}}\frac{(k_\perp^2)^{-2\epsilon}}{\Gamma(1-2\epsilon)}\,d\theta_{12}^2\,d\theta^2_{13}\, d\phi\,dz_i\,dz_j\,z_i^{1-2\epsilon}\,z_j^{1-2\epsilon}\,z_k^{1-2\epsilon}(\theta_{12}^2)^{-\epsilon}(\theta_{13}^2)^{-\epsilon}\sin^{-2\epsilon}\phi\,,
\end{align}
for phase space coordinates formed from pairwise angles $\theta_{12}^2,\theta_{13}^2$, azimuthal angle $\phi$, and energy fractions $z_i$, with $z_1+z_2+z_3=1$.  We have rescaled out the jet energy $E^2$ so that only the low-scale $k_\perp^2$ appears here.  By the law of cosines, we have the relationship that
\begin{align}
\theta_{23}^2 = \theta_{12}^2+\theta_{13}^2-2\theta_{12}\theta_{13}\cos\phi\,.
\end{align}
However, in most expressions we will retain $\theta_{23}^2$ with implicit dependence on azimuthal angle $\phi$ for brevity.

For the flavor transition $g\to q$, there is only a single final state particle configuration that can contribute, namely, the process $g\to gq\bar q$, which contributes to two distinct color channels as defined by the emission structure of the final state gluon.  The squared matrix element can then be expressed as
\begin{align}
|{\cal M}_{g\to g_1q_2 \bar q_3}|^2 = \left(
\frac{\alpha_s}{2\pi}
\right)^2\frac{(4\pi)^4(\mu_0^2)^{2\epsilon}}{(z_1z_2\theta_{12}^2+z_1z_3\theta_{13}^2+z_2z_3\theta_{23}^2)^2}\left[
C_FT_R P^\text{ab}_{g\to \bar g_1q_2 \bar q_3}+C_AT_R P^\text{nab}_{g\to \bar g_1q_2 \bar q_3}
\right]\,,
\end{align}
where $T_R = 1/2$ and $C_F=4/3$ and $C_A=3$ are the fundamental and adjoint quadratic Casimirs of SU(3) flavor of QCD.  The corresponding splitting functions are separated in Abelian (ab) or non-Abelian (nab) contributions and take the form \cite{Catani:1999ss,Campbell:1997hg}
\begin{align}
P^\text{ab}_{g\to g_1q_2 \bar q_3} &= -2-(1-\epsilon)z_2z_3\theta_{23}^2\left(
\frac{1}{z_1z_2\theta_{12}^2}+\frac{1}{z_1z_3\theta_{13}^2}
\right)\\
&\hspace{1cm}+2\frac{(z_1z_2\theta_{12}^2+z_1z_3\theta_{13}^2+z_2z_3\theta_{23}^2)^2}{z_1^2z_2z_3\theta_{12}^2\theta_{13}^2}\left(
1+z_1^2-\frac{z_1+2z_2z_3}{1-\epsilon}
\right)\nonumber\\
&\hspace{1cm} - \frac{z_1z_2\theta_{12}^2+z_1z_3\theta_{13}^2+z_2z_3\theta_{23}^2}{z_1z_2\theta_{12}^2}\left(
1+2z_1+\epsilon-2\frac{z_1+z_2}{1-\epsilon}
\right)\nonumber\\
&\hspace{1cm} - \frac{z_1z_2\theta_{12}^2+z_1z_3\theta_{13}^2+z_2z_3\theta_{23}^2}{z_1z_3\theta_{13}^2}\left(
1+2z_1+\epsilon-2\frac{z_1+z_3}{1-\epsilon}
\right)\nonumber\,,\\
P^\text{nab}_{g\to g_1q_2 \bar q_3} &=-\frac{1}{2\theta_{23}^4}\left(
\frac{2z_1}{z_2+z_3}(\theta_{13}^2-\theta_{12}^2)+\frac{z_2-z_3}{z_2+z_3}\theta_{23}^2
\right)^2-\frac{1}{2}+\epsilon\\
&\hspace{1cm}-\frac{(z_1z_2\theta_{12}^2+z_1z_3\theta_{13}^2+z_2z_3\theta_{23}^2)^2}{z_1^2z_2z_3\theta_{12}^2\theta_{13}^2}\left(
1+z_1^2-\frac{z_1+2z_2z_3}{1-\epsilon}
\right)
\nonumber\\
&\hspace{1cm}
+\frac{z_1z_2\theta_{12}^2+z_1z_3\theta_{13}^2+z_2z_3\theta_{23}^2}{z_2z_3\theta_{23}^2}\left(
\frac{1+z_1^3}{z_1(1-z_1)}+\frac{z_1(z_3-z_2)^2-2z_2z_3(1+z_1)}{(1-\epsilon)z_1(1-z_1)}
\right)
\nonumber\\
&\hspace{1cm}
+\frac{z_1z_2\theta_{12}^2+z_1z_3\theta_{13}^2+z_2z_3\theta_{23}^2}{2z_1z_3\theta_{13}^2}(1-z_2)\left(
1+\frac{1}{z_1(1-z_1)}-\frac{2z_2(1-z_2)}{(1-\epsilon)z_1(1-z_1)}
\right)
\nonumber\\
&\hspace{1cm}
+\frac{z_1z_2\theta_{12}^2+z_1z_3\theta_{13}^2+z_2z_3\theta_{23}^2}{2z_1z_2\theta_{12}^2}(1-z_3)\left(
1+\frac{1}{z_1(1-z_1)}-\frac{2z_3(1-z_3)}{(1-\epsilon)z_1(1-z_1)}
\right)\nonumber\\
&
\hspace{1cm}+\frac{(z_1z_2\theta_{12}^2+z_1z_3\theta_{13}^2+z_2z_3\theta_{23}^2)^2}{2z_1z_2z_3\theta_{13}^2\theta_{23}^2}\left(
\frac{(1-z_1)^3-z_1^3}{z_1(1-z_1)}-\frac{2z_3(1-z_3-2z_1z_2)}{(1-\epsilon)z_1(1-z_1)}
\right)
\nonumber\\
&
\hspace{1cm}+\frac{(z_1z_2\theta_{12}^2+z_1z_3\theta_{13}^2+z_2z_3\theta_{23}^2)^2}{2z_1z_2z_3\theta_{12}^2\theta_{23}^2}\left(
\frac{(1-z_1)^3-z_1^3}{z_1(1-z_1)}-\frac{2z_2(1-z_2-2z_1z_3)}{(1-\epsilon)z_1(1-z_1)}
\right)
\nonumber\,.
\end{align}

Finally, we need the measurement function.  For final state particles labeled as $g_1q_2 \bar q_3$, the measurement function takes the form
\begin{align}
\hspace{-0.225cm}\Theta_\text{WTA}& = \Theta(\min[z_1^2,z_3^2]\theta_{13}^2-\min[z_1^2,z_2^2]\theta_{12}^2)\Theta(\min[z_2^2,z_3^2]\theta_{23}^2-\min[z_1^2,z_2^2]\theta_{12}^2)\Theta\left(\frac{1}{2}-z_3\right)\\
&\hspace{1cm}\times\left[
\Theta(k_\perp^2-z_2^2\theta_{12}^2E^2)\Theta(z_1-z_2)\Theta(z_3^2\theta_{13}^2E^2-k_\perp^2)+\Theta(z_2-z_1)\Theta(z_3^2\theta_{23}^2E^2-k_\perp^2)
\right]\nonumber\\
&+\Theta(\min[z_1^2,z_2^2]\theta_{12}^2-\min[z_1^2,z_3^2]\theta_{13}^2)\Theta(\min[z_2^2,z_3^2]\theta_{23}^2-\min[z_1^2,z_3^2]\theta_{13}^2)\Theta\left(z_2-\frac{1}{2}\right)\nonumber\\
&\hspace{1cm}\times\left[
\Theta(z_1-z_3)\Theta((1-z_2)^2\theta_{12}^2E^2-k_\perp^2)+\Theta(z_3-z_1)\Theta((1-z_2)^2\theta_{23}^2E^2-k_\perp^2)
\right]\nonumber\\
&+\Theta(\min[z_1^2,z_3^2]\theta_{13}^2-z_3^2\theta_{23}^2)\Theta(\min[z_1^2,z_2^2]\theta_{12}^2-z_3^2\theta_{23}^2)\Theta\left(\frac{1}{2}-z_1\right)\Theta(z_2-z_3)\Theta(z_3^2\theta_{23}^2E^2-k_\perp^2)\nonumber\,.
\end{align}
In the first two lines, particles 1 and 2 are closest according to the $k_T$ metric.  For this configuration to be labeled as quark $q$, particle 3 ($\bar q$) must be both relatively soft ($z_3 < 1/2$) and particle 3 cannot be clustered within $k_\perp^2$ of particles 1 and 2 (second line).  Note also that if the gluon is more energetic than the quark, $z_1>z_2$, the $gq$ splitting must be unresolved and have a relative transverse momentum less than $k_\perp^2$.  If the quark is more energetic than the gluon, $z_2>z_1$, then there is no restriction on their relative transverse momentum.  In lines three and four, particles 1 and 3 are clustered together first, and so particle 2 (the quark) must take most of the jet energy ($z_2 > 1/2$).  On the fourth line, particles 1 and 3 cannot be clustered within $k_\perp^2$ of particle 2.  On the final line, particles 2 and 3 are clustered together first.  The gluon must therefore be relatively soft ($z_1 < 1/2$), the quark relatively hard ($z_2 > z_3$) and particles 2 and 3 must be resolved, outside of the scale $k_\perp^2$.

To proceed with the calculation, we first rescale all angles by $\theta_{13}^2$, $\theta^2 \to \theta_{13}^2\theta^2$.  Because the splitting functions are homogeneous in overall angle, this produces a simple integral over $\theta_{13}^2$ that can be evaluated.  With this change of variables, the bare jet function takes the form
\begin{align}
J^{(2,\text{bare})}_{g\to q}(\mu_0^2,k_\perp^2) &=\left(
\frac{\alpha_s}{2\pi}
\right)^2\frac{4^{-\epsilon}}{\pi}\left(\frac{\mu_0^2}{k_\perp^2}\right)^{2\epsilon}\frac{1}{\Gamma(1-2\epsilon)}\\
&\hspace{1cm}
\times\int d\theta_{12}^2\,\frac{d\theta^2_{13}}{(\theta^2_{13})^{1+2\epsilon}}\, d\phi\,dz_i\,dz_j\,z_i^{1-2\epsilon}\,z_j^{1-2\epsilon}\,z_k^{1-2\epsilon}(\theta_{12}^2)^{-\epsilon}\sin^{-2\epsilon}\phi\left[\cdot\right]_{\theta_{13}^2\to 1}\nonumber\,,
\end{align}
where $\left[\cdot\right]_{\theta_{13}^2\to 1}$ denotes both the measurement function and the matrix element of the rescale coordinates, setting $\theta_{13}^2\to 1$.  The measurement function with this rescaling becomes
\begin{align}
\Theta_\text{WTA}& = \Theta(\min[z_1^2,z_3^2]-\min[z_1^2,z_2^2]\theta_{12}^2)\Theta(\min[z_2^2,z_3^2]\theta_{23}^2-\min[z_1^2,z_2^2]\theta_{12}^2)\Theta\left(\frac{1}{2}-z_3\right)\nonumber\\
&\hspace{1cm}\times\left[
\Theta(1-z_2^2\theta_{12}^2\theta_{13}^2)\Theta(z_1-z_2)\Theta(z_3^2\theta_{13}^2-1)+\Theta(z_2-z_1)\Theta(z_3^2\theta_{23}^2\theta_{13}^2-1)
\right]\\
&+\Theta(\min[z_1^2,z_2^2]\theta_{12}^2-\min[z_1^2,z_3^2])\Theta(\min[z_2^2,z_3^2]\theta_{23}^2-\min[z_1^2,z_3^2])\Theta\left(z_2-\frac{1}{2}\right)\nonumber\\
&\hspace{1cm}\times\left[
\Theta(z_1-z_3)\Theta((1-z_2)^2\theta_{12}^2\theta_{13}^2-1)+\Theta(z_3-z_1)\Theta((1-z_2)^2\theta_{23}^2\theta_{13}^2-1)
\right]\nonumber\\
&+\Theta(\min[z_1^2,z_3^2]-z_3^2\theta_{23}^2)\Theta(\min[z_1^2,z_2^2]\theta_{12}^2-z_3^2\theta_{23}^2)\Theta\left(\frac{1}{2}-z_1\right)\Theta(z_2-z_3)\Theta(z_3^2\theta_{23}^2\theta_{13}^2-1)\nonumber\,,
\end{align}
where we have also rescaled angles by the ratio $k_\perp^2/E^2$.  The jet function can then be integrated over $\theta_{13}^2$ to produce
\begin{align}
J^{(2,\text{bare})}_{g\to q}(\mu_0^2,k_\perp^2) &=\left(
\frac{\alpha_s}{2\pi}
\right)^2\frac{4^{-\epsilon}}{\pi}\left(\frac{\mu_0^2}{k_\perp^2}\right)^{2\epsilon}\frac{1}{\Gamma(1-2\epsilon)}\frac{1}{2\epsilon}\\
&\hspace{1cm}
\times\int d\theta_{12}^2\, d\phi\,dz_1\,dz_2\,z_1^{1-2\epsilon}\,z_2^{1-2\epsilon}\,z_3^{1-2\epsilon}(\theta_{12}^2)^{-\epsilon}\sin^{-2\epsilon}\phi\left[\cdot\right]_{\theta_{13}^2\to 1}\nonumber\,,
\end{align}
and the measurement function is then
\begin{align}
\Theta_\text{WTA}& = \Theta(\min[z_1^2,z_3^2]-\min[z_1^2,z_2^2]\theta_{12}^2)\Theta(\min[z_2^2,z_3^2]\theta_{23}^2-\min[z_1^2,z_2^2]\theta_{12}^2)\Theta\left(\frac{1}{2}-z_3\right)\\
&\hspace{-1cm}\times\left[
\left((z_3^2)^{2\epsilon}-(z_2^2\theta_{12}^2)^{2\epsilon}\right)\Theta(z_1-z_2)\Theta(z_3^2 - z_2^2\theta_{12}^2)+\left(z_3^2\theta_{23}^2\right)^{2\epsilon}\Theta(z_2-z_1)
\right]\nonumber\\
&+\Theta(\min[z_1^2,z_2^2]\theta_{12}^2-\min[z_1^2,z_3^2])\Theta(\min[z_2^2,z_3^2]\theta_{23}^2-\min[z_1^2,z_3^2])\Theta\left(z_2-\frac{1}{2}\right)\nonumber\\
&\hspace{1cm}\times\left[
((1-z_2)^2\theta_{12}^2)^{2\epsilon}\Theta(z_1-z_3)+((1-z_2)^2\theta_{23}^2)^{2\epsilon}\Theta(z_3-z_1)
\right]\nonumber\\
&+(z_3^2\theta_{23}^2)^{2\epsilon}\Theta(\min[z_1^2,z_3^2]-z_3^2\theta_{23}^2)\Theta(\min[z_1^2,z_2^2]\theta_{12}^2-z_3^2\theta_{23}^2)\Theta\left(\frac{1}{2}-z_1\right)\Theta(z_2-z_3)\nonumber\,.
\end{align}

We can now extract the singular contributions to the jet function by explicit expansion in the soft and/or collinear limits.  We will consider the Abelian and non-Abelian contributions separately as they have different divergences.

\subsection{Singular Abelian Contribution}

For the Abelian, $C_F$, splitting function, there are three singular limits: the soft gluon limit, $z_1\to 0$, the collinear limit $\theta_{12}^2\to 0$ and the collinear limit $\theta_{13}^2\to 0$, which, after rescaling by $\theta_{13}^2$, is manifest as the limit $\theta_{12}^2\to \infty$.  We consider each of these limits in turn.

\subsubsection{Soft Limit}

In the soft $z_1\to 0$ limit, the measurement function reduces to 
\begin{align}
\Theta_\text{WTA}^\text{soft}& \to\left(z_3^2\theta_{23}^2\right)^{2\epsilon} \Theta(1-\theta_{12}^2)\Theta\left(\frac{1}{2}-z_3\right)+((1-z_2)^2\theta_{23}^2)^{2\epsilon}\Theta(\theta_{12}^2-1)\Theta\left(z_2-\frac{1}{2}\right)\,,
\end{align}
while the Abelian matrix element is
\begin{align}
|{\cal M}^\text{ab,soft}_{g\to g_1q_2 \bar q_3}|^2 &\to\frac{2C_F T_R}{z_1^2z_2z_3\theta_{12}^2}
\left(
1-\frac{2z_2z_3}{1-\epsilon}
\right)\,.
\end{align}
In this expression, we have pulled out dependence on scales and the strong coupling.  The contribution to the jet function is then
\begin{align}
J^{(2,\text{bare,soft})}_{g\to q,\text{ab}}(\mu_0^2,k_\perp^2) &=\left(
\frac{\alpha_s}{2\pi}
\right)^2\frac{4^{-\epsilon}}{\pi}\left(\frac{\mu_0^2}{k_\perp^2}\right)^{2\epsilon}\frac{1}{\Gamma(1-2\epsilon)}\frac{1}{2\epsilon}\\
&\hspace{1cm}
\times\int d\theta_{12}^2\, d\phi\,dz_1\,dz_2\,z_1^{1-2\epsilon}\,z_2^{1-2\epsilon}\,z_3^{1-2\epsilon}(\theta_{12}^2)^{-\epsilon}\sin^{-2\epsilon}\phi\,|{\cal M}^\text{ab,soft}_{g\to g_1q_2 \bar q_3}|^2\Theta_\text{WTA}^\text{soft}\nonumber\\
&=\left(
\frac{\alpha_s}{2\pi}
\right)^2\left(\frac{\mu_0^2}{k_\perp^2}\right)^{2\epsilon}\frac{C_F T_R}{\epsilon^2}\nonumber\\
&
\hspace{1cm}\times\left[
\frac{1}{3}\frac{1}{\epsilon} + \frac{1}{18}-\frac{4}{3}\log 2+\epsilon\left(
-\frac{61}{54}-\frac{11}{9}\log 2-\frac{4}{3}\log^2 2-\frac{\pi^2}{18}
\right)+\cdots
\right]\nonumber\,,
\end{align}
where the ellipses are higher order in $\epsilon$ and not needed for extraction of anomalous dimensions.  To perform the $\epsilon$ expansion of the result of the analytic integrals here and everywhere exact results are quoted, we used the HypEXP 2.0 Mathematica package \cite{Huber:2005yg,Huber:2007dx}.

\subsubsection{Collinear Limits}

There are two collinear limits, corresponding to $\theta_{12}^2\to 0$ and $\theta_{12}^2\to \infty$.  The measurement function of the first collinear limit is
\begin{align}
\Theta_\text{WTA}^0& \to \left[z_3^{4\epsilon} - z_2^{4\epsilon}(\theta_{12}^2)^{2\epsilon}\Theta(z_1-z_2)\right]\Theta\left(\frac{1}{2}-z_3\right)\Theta(1-\theta_{12}^2)\,.
\end{align}
This can be further separated into two constraints that can be evaluated individually.  The first constraint is identical to that in the collinear limit of the soft calculation above, and so to avoid double counting, the relevant collinear matrix element must lack a soft limit:
\begin{align}
|{\cal M}_{g\to g_1q_2 \bar q_3}^{\text{ab},0}|^2&\to \frac{C_F T_R}{z_3^2(1-z_3)^2}
\frac{1}{\theta_{12}^2}\left[
-(1-\epsilon)
\frac{z_3}{z_1}+2\frac{z_3(1-z_3)^2}{z_1z_2}\left(
z_1-\frac{1}{1-\epsilon}
\right)\right.\\
&\hspace{4cm} \left.- \frac{z_3(1-z_3)}{z_1z_2}\left(
1+2z_1+\epsilon-2\frac{z_1+z_2}{1-\epsilon}
\right)
\right]\nonumber\,.
\end{align}
Its contribution to the jet function is then
\begin{align}
J^{(2,\text{bare}),0}_{g\to q,\text{ab}}(\mu_0^2,k_\perp^2) &\supset-\left(
\frac{\alpha_s}{2\pi}
\right)^2\left(\frac{\mu_0^2}{k_\perp^2}\right)^{2\epsilon}\frac{C_F T_R}{\Gamma(1-\epsilon)^2}\frac{1}{2\epsilon^2}\int dz_1\,dz_2\,z_1^{1-2\epsilon}\,z_2^{1-2\epsilon}\,z_3^{1+2\epsilon}|{\cal M}_{g\to g_1q_2 \bar q_3}^{\text{ab},0}|^2\Theta_\text{WTA}^0
\nonumber\\
&=\left(
\frac{\alpha_s}{2\pi}
\right)^2\left(\frac{\mu_0^2}{k_\perp^2}\right)^{2\epsilon}\frac{C_F T_R}{\epsilon^2}\left[
\frac{5}{12}+\epsilon\left(
\frac{385}{144}-\frac{5}{2}\log 2
\right)+\cdots
\right]\,.
\end{align}
For the second part of the measurement function, the soft limit is already regulated and so the relevant matrix element is
\begin{align}
|{\cal M}_{g\to g_1q_2 \bar q_3}^{\text{ab},0}|^2&\to \frac{C_F T_R}{z_3^2(1-z_3)^2}
\frac{1}{\theta_{12}^2}\left[
-(1-\epsilon)
\frac{z_3}{z_1}+2\frac{z_3(1-z_3)^2}{z_1^2z_2}\left(1+
z_1^2-\frac{z_1+2z_2z_3}{1-\epsilon}
\right)\right.\\
&\hspace{4cm} \left.- \frac{z_3(1-z_3)}{z_1z_2}\left(
1+2z_1+\epsilon-2\frac{z_1+z_2}{1-\epsilon}
\right)
\right]\nonumber\,.
\end{align}
The contribution to the jet function is then
\begin{align}
J^{(2,\text{bare}),0}_{g\to q,\text{ab}}(\mu_0^2,k_\perp^2) &\supset-\left(
\frac{\alpha_s}{2\pi}
\right)^2\left(\frac{\mu_0^2}{k_\perp^2}\right)^{2\epsilon}\frac{C_F T_R}{\Gamma(1-\epsilon)^2}\frac{1}{2\epsilon^2}\int dz_1\,dz_2\,z_1^{1-2\epsilon}\,z_2^{1+2\epsilon}\,z_3^{1-2\epsilon}|{\cal M}_{g\to g_1q_2 \bar q_3}^{\text{ab},0}|^2\Theta_\text{WTA}^0\nonumber\\
&=\left(
\frac{\alpha_s}{2\pi}
\right)^2\left(\frac{\mu_0^2}{k_\perp^2}\right)^{2\epsilon}\frac{C_F T_R}{\epsilon^2}\left[
\frac{5}{48}-\frac{1}{3}\log 2\right.\\
&\hspace{4cm}\left.+\,\epsilon\left(
\frac{199}{576}-\frac{67}{72}\log 2 - \frac{2}{3}\log^2 2+\frac{\pi^2}{18}
\right)+\cdots
\right]\,.
\nonumber
\end{align}

Moving to the $\theta_{12}^2\to \infty$ limit, the measurement function is 
\begin{align}
\Theta_\text{WTA}^\infty& \to((1-z_2)^2\theta_{12}^2)^{2\epsilon} \Theta\left(z_2-\frac{1}{2}\right)\Theta(\theta_{12}^2-1)\,.
\end{align}
The matrix element in this case reduces to
\begin{align}
|{\cal M}_{g\to g_1q_2 \bar q_3}^{\text{ab},\infty}|^2 &\to \frac{C_FT_R}{z_2^2(1-z_2)^2}
\frac{1}{\theta_{12}^2}\left[
-(1-\epsilon)
\frac{z_2}{z_1}+2\frac{z_2(1-z_2)^2}{z_1z_3}\left(
z_1-\frac{1}{1-\epsilon}
\right)\right.\\
&\hspace{4cm} \left.- \frac{z_2(1-z_2)}{z_1z_3}\left(
1+2z_1+\epsilon-2\frac{z_1+z_3}{1-\epsilon}
\right)
\right]\nonumber\,,
\end{align}
where, as above, we explicitly subtract the soft limit within the matrix element to avoid double counting.  The contribution to the jet function from this limit is then
\begin{align}
J^{(2,\text{bare}),\infty}_{g\to q,\text{ab}}(\mu_0^2,k_\perp^2) &=-\left(
\frac{\alpha_s}{2\pi}
\right)^2\left(\frac{\mu_0^2}{k_\perp^2}\right)^{2\epsilon}\frac{C_F T_R}{\Gamma(1-\epsilon)^2}\frac{1}{2\epsilon^2}\int dz_1\,dz_2\,z_1^{1-2\epsilon}\,z_2^{1-2\epsilon}\,z_3^{1-2\epsilon}|{\cal M}_{g\to g_1q_2 \bar q_3}^{\text{ab},\infty}|^2\Theta_\text{WTA}^\infty\nonumber\\
&=\left(
\frac{\alpha_s}{2\pi}
\right)^2\left(\frac{\mu_0^2}{k_\perp^2}\right)^{2\epsilon}\frac{C_F T_R}{\epsilon^2}\left[
\frac{5}{12}+\epsilon\left(
\frac{379}{144}-\frac{5}{6}\log 2
\right)+\cdots
\right]\,.
\end{align}

\subsubsection{Remaining Numerical Integral}

With the divergent contributions explicitly isolated and calculated, the remaining contribution at $1/\epsilon$ order can be expressed as a finite, numerical integral.  That is, from the expression of the jet function master formula after integrating over overall angular size, we can set $\epsilon = 0$ everywhere, except exactly at the pole.  That is, we have
\begin{align}
J^{(2,\text{bare,num})}_{g\to q}(\mu_0^2,k_\perp^2) &=\left(
\frac{\alpha_s}{2\pi}
\right)^2\left(\frac{\mu_0^2}{k_\perp^2}\right)^{2\epsilon}\frac{1}{2\pi \epsilon}\int d\theta_{12}^2\, d\phi\,dz_1\,dz_2\,z_1\,z_2\,z_3\left[\cdot\right]_{\theta_{13}^2\to 1,\epsilon\to 0}\,,
\end{align}
where $\left[\cdot\right]_{\theta_{13}^2\to 1,\epsilon\to 0}$ is the appropriate matrix element and measurement functions with $\epsilon = 0$ and subtracted of any and all soft and collinear limits.  In this case, the complete measurement function reduces to, for example,
\begin{align}\label{eq:wta0eps}
\Theta_\text{WTA}^{\epsilon\to 0}& = \Theta(\min[z_1^2,z_3^2]-\min[z_1^2,z_2^2]\theta_{12}^2)\Theta(\min[z_2^2,z_3^2]\theta_{23}^2-\min[z_1^2,z_2^2]\theta_{12}^2)\Theta\left(\frac{1}{2}-z_3\right)\Theta(z_2-z_1)\nonumber\\
&+\Theta(\min[z_1^2,z_2^2]\theta_{12}^2-\min[z_1^2,z_3^2])\Theta(\min[z_2^2,z_3^2]\theta_{23}^2-\min[z_1^2,z_3^2])\Theta\left(z_2-\frac{1}{2}\right)\nonumber\\
&+\Theta(\min[z_1^2,z_3^2]-z_3^2\theta_{23}^2)\Theta(\min[z_1^2,z_2^2]\theta_{12}^2-z_3^2\theta_{23}^2)\Theta\left(\frac{1}{2}-z_1\right)\Theta(z_2-z_3)\,.
\end{align}
To perform the integration, we use the implementation of VEGAS \cite{Lepage:1977sw,Lepage:1980dq} in Cuba 4.2 \cite{Hahn:2004fe}.  In this Abelian channel, we find the result of this numerical integration to be
\begin{align}
J^{(2,\text{bare,num})}_{g\to q,\text{ab}}(\mu_0^2,k_\perp^2) &=\left(
\frac{\alpha_s}{2\pi}
\right)^2C_F T_R\left(\frac{\mu_0^2}{k_\perp^2}\right)^{2\epsilon}\frac{-0.6378}{\epsilon}\,,
\end{align}
with error at the order of the last quoted digit.

\subsubsection{Total Real Emission Contribution}

Summing the soft, collinear, and numerical contributions, we find the contribution to the bare jet function of
\begin{align}\label{eq:ab2loopgq}
J^{(2,\text{bare}),\text{sub}}_{g\to q,\text{ab}}(\mu_0^2,k_\perp^2) &=\left(
\frac{\alpha_s}{2\pi}
\right)^2C_F T_R\left(\frac{\mu_0^2}{k_\perp^2}\right)^{2\epsilon}\left[\frac{1}{3\epsilon^3}
+\frac{1}{\epsilon^2}\left(\frac{143}{144}-\frac{5}{3}\log 2\right)\right.\\
&\hspace{4cm}\left.+\frac{1}{\epsilon}\left(
\frac{7813}{1728}-\frac{395}{72}\log 2-2\log^2 2-0.6378
\right)+\cdots
\right]\nonumber\,.
\end{align}

\subsection{Singular Non-Abelian Contribution}

We now move to calculation of the singular limits of the non-Abelian, $C_A$, contribution.  There are also three singular limits: a soft gluon, $z_1\to 0$, collinear quark-anti-quark $\theta_{23}^2\to 0$, and the soft $q\bar q$ limit, $z_2,z_3\to 0$, though we will see that dealing with each will be rather different than in the Abelian channel.

\subsubsection{Soft Gluon Limit}

In the limit that $z_1\to 0$, the measurement function simplifies to
\begin{align}
\Theta_\text{WTA}^\text{1~soft}& \to\left(z_3^2\theta_{23}^2\right)^{2\epsilon} \Theta\left(\frac{1}{2}-z_3\right)\,.
\end{align}
The corresponding matrix element in this limit takes the form 
\begin{align}
|{\cal M}^\text{nab,1~soft}_{g\to g_1q_2 \bar q_3}|^2 &\to\frac{C_A T_R}{z_1^2z_2z_3\theta_{12}^2\theta_{23}^2}
(1+\theta_{12}^2-\theta_{23}^2)\left(
1-\frac{2z_2z_3}{1-\epsilon}
\right)\,.
\end{align}
Again, we have written the matrix element in terms of relative angles, but the azimuthal angle can be introduced with the law of cosines.  The contribution to the jet function from this soft limit is then
\begin{align}
J^{(2,\text{bare, 1~soft})}_{g\to q,\text{nab}}(\mu_0^2,k_\perp^2) &=\left(
\frac{\alpha_s}{2\pi}
\right)^2\frac{4^{-\epsilon}}{\pi}\left(\frac{\mu_0^2}{k_\perp^2}\right)^{2\epsilon}\frac{1}{\Gamma(1-2\epsilon)}\frac{1}{2\epsilon}\\
&\hspace{1cm}
\times\int d\theta_{12}^2\, d\phi\,dz_1\,dz_2\,z_1^{1-2\epsilon}\,z_2^{1-2\epsilon}\,z_3^{1-2\epsilon}(\theta_{12}^2)^{-\epsilon}\sin^{-2\epsilon}\phi\,|{\cal M}^\text{nab,1~soft}_{g\to g_1q_2 \bar q_3}|^2\Theta_\text{WTA}^\text{1~soft}\nonumber\\
&=\left(
\frac{\alpha_s}{2\pi}
\right)^2\left(\frac{\mu_0^2}{k_\perp^2}\right)^{2\epsilon}\frac{C_A T_R}{\epsilon^2}\nonumber\\
&\hspace{1cm}
\times\left[
-\frac{1}{6\epsilon}+\frac{1}{72}+\log 2+\epsilon\left(
\frac{23}{54}-\frac{\pi^2}{12}+\frac{11}{12}\log 2+\log^22
\right)+\cdots
\right]\nonumber\,.
\end{align}

\subsubsection{Collinear Quark Limit}

In the limit that the $q\bar q$ pair becomes collinear, $\theta_{23}^2\to 0$, the measurement function reduces to
\begin{align}
\Theta_\text{WTA}^\text{23~coll}& \to (z_3^2\theta_{23}^2)^{2\epsilon}\Theta(1-\theta_{23}^2)\Theta\left(\frac{1}{2}-z_1\right)\Theta(z_2-z_3)\,.
\end{align}
The matrix element in this limit reduces to
\begin{align}
|{\cal M}^\text{nab,2,3~coll}_{g\to g_1q_2 \bar q_3}|^2 &\to \frac{C_A T_R}{z_1^2z_2z_3(1-z_1)^2\theta_{23}^2}\left[
-\frac{8z_1^2z_2z_3}{(1-z_1)^2}\cos^2\phi+
1+z_1^3+\frac{z_1(z_3-z_2)^2-2z_2z_3(1+z_1)}{1-\epsilon}
\right.
\nonumber\\
&
\hspace{1cm}+\frac{1-z_1}{2}\left(
(1-z_1)^3-z_1^3-\frac{2z_3(1-z_3-2z_1z_2)}{1-\epsilon}
\right)
\nonumber\\
&
\hspace{1cm}\left.+\frac{1-z_1}{2}\left(
(1-z_1)^3-z_1^3-\frac{2z_2(1-z_2-2z_1z_3)}{1-\epsilon}
\right)
\nonumber
\right]\,.
\end{align}
This still has a soft limit as $z_1\to 0$, which we already calculated above, so we have to explicitly subtract the soft and collinear limit.  The measurement function in this limit is
\begin{align}
\Theta_\text{WTA}^\text{1~soft, 2,3~coll}& \to\left(z_3^2\theta_{23}^2\right)^{2\epsilon} \Theta(1-\theta_{23}^2)\Theta\left(\frac{1}{2}-z_3\right)\,,
\end{align}
and the corresponding matrix element is
\begin{align}
|{\cal M}^\text{nab,1~soft,2,3~coll}_{g\to g_1q_2 \bar q_3}|^2 &\to\frac{2C_A T_R}{z_1^2z_2z_3\theta_{23}^2}
\left(
1-\frac{2z_2z_3}{1-\epsilon}
\right)\,.
\end{align}
The collinear $q\bar q$ contribution to the jet function is therefore
\begin{align}
J^{(2,\text{bare, 23~coll})}_{g\to q,\text{nab}}(\mu_0^2,k_\perp^2) &=\left(
\frac{\alpha_s}{2\pi}
\right)^2\frac{4^{-\epsilon}}{\pi}\left(\frac{\mu_0^2}{k_\perp^2}\right)^{2\epsilon}\frac{1}{\Gamma(1-2\epsilon)}\frac{1}{2\epsilon}\\
&\hspace{1cm}
\times\int d\theta_{12}^2\, d\phi\,dz_1\,dz_2\,z_1^{1-2\epsilon}\,z_2^{1-2\epsilon}\,z_3^{1-2\epsilon}(\theta_{12}^2)^{-\epsilon}\sin^{-2\epsilon}\phi\nonumber\\
&\hspace{2cm}\times\left(
|{\cal M}^\text{nab,2,3~coll}_{g\to g_1q_2 \bar q_3}|^2\Theta_\text{WTA}^\text{23~coll} - |{\cal M}^\text{nab,1~soft,2,3~coll}_{g\to g_1q_2 \bar q_3}|^2 \Theta_\text{WTA}^\text{1~soft, 2,3~coll}
\right)\nonumber\\
&=\left(
\frac{\alpha_s}{2\pi}
\right)^2\left(\frac{\mu_0^2}{k_\perp^2}\right)^{2\epsilon}\frac{C_A T_R}{\epsilon^2}
\left[
-\frac{11}{72}-\frac{\log 2}{3}+\epsilon\left(
-\frac{199}{216}-\frac{23}{36}\log 2-\log^2 2+\frac{\pi^2}{18}
\right)
\right]\nonumber\,.
\end{align}

\subsubsection{Soft Quark Limit}

The final singular limit present in the non-Abelian matrix element is the $q\bar q$ soft limit, $z_2,z_3\to 0$.  In this limit the measurement function is
\begin{align}
\Theta_\text{WTA}^\text{2,3~soft}& \to\left(z_3^{4\epsilon}-z_2^{4\epsilon}\theta_{12}^{4\epsilon}\right) \Theta(z_3^2-z_2^2\theta_{12}^2)\Theta(\min[z_2^2,z_3^2]\theta_{23}^2-z_2^2\theta_{12}^2)\,,
\end{align}
and the matrix element reduces to
\begin{align}
|{\cal M}^\text{nab,2,3~soft}_{g\to g_1q_2 \bar q_3}|^2 &\to \frac{2C_A T_R}{(z_2\theta_{12}^2+z_3)^2} \left[
-\frac{1}{\theta_{23}^4}\left(
\frac{1-\theta_{12}^2}{z_2+z_3}
\right)^2
+\frac{z_2\theta_{12}^2+z_3}{z_2z_3\theta_{23}^2}
\frac{1}{z_2+z_3}
\right]\,.
\end{align}
What is especially interesting about this limit is that there is a rapidity divergence, unregulated by dimensional regularization.  Because we have already integrated over the overall angle of the splitting, this rapidity divergence manifests itself as an unregulated energy divergence:
\begin{align}
\int d\Phi_3\, |{\cal M}^\text{nab,2,3~soft}_{g\to g_1q_2 \bar q_3}|^2\, \Theta_\text{WTA}^\text{2,3~soft} \sim \int_0^\infty \frac{dz}{z}\,,
\end{align}
where $z$ is the energy scale of the $q\bar q$ pair.  From the perspective of the factorization theorem, this must not contribute because soft divergences cannot affect collinear evolution.

To resolve this rapidity divergence, we introduce a rapidity regulator \cite{Chiu:2011qc,Chiu:2012ir}, which explicitly deforms the energy distribution of the soft gluon to quark--anti-quark splitting away from the mass shell.  Here, the regulator we apply is to rescale the sum of the energy of the $q\bar q$ pair, $E_g$, by a factor
\begin{align}
E_g \to (z_2+z_3)^{-\eta}\,.
\end{align}
This modifies the corresponding energy integral to 
\begin{align}
\int d\Phi_3\, |{\cal M}^\text{nab,2,3~soft}_{g\to g_1q_2 \bar q_3}|^2\, \Theta_\text{WTA}^\text{2,3~soft} \sim \int_0^\infty \frac{dz}{z^{1+\eta}} = 0\,,
\end{align}
a scaleless integral and therefore equal to 0.  Then, as rapidity regularization cannot affect dimensional regularization, we take $\eta \to 0$, which has no effect because the contribution already vanished.  Further, the measurement function is suppressed in the regulator $\epsilon$ by the difference $z_3^{4\epsilon}-z_2^{4\epsilon}\theta_{12}^{4\epsilon}\propto \epsilon$, rendering any possible divergences in $\epsilon$ finite.  Thus, the soft $q\bar q$ limit does not contribute to singularities in the bare jet function at all and correspondingly does not affect anomalous dimensions, again, as required by the structure of the factorization theorem.

The vanishing of this soft limit indeed can be interpreted as a cancelation of infrared (from the jet function) and ultraviolet (from the hard function) divergences.  What this means is that this factorization theorem forces this soft limit to be associated with infrared flavor of the hard, initial parton, even though the net flavor according to the phase space constraints is determined by the soft quark itself.  This only happens because of this leading-power factorization theorem, that the scales of the jet and hard functions have been pulled effectively infinitely far from one another.  In a practical implementation of the WTA flavor algorithm, with non-zero and finite ratios between scales and finite phase space volumes, perfect cancellation of soft divergences is not automatic, and must be carefully addressed when matching the factorization theorem to fixed-order.  Therefore, the virtual and real soft quark divergences are necessarily perfectly canceled between the hard and jet functions with this prescription, rendering all divergences that require renormalization of hard collinear origin.

\subsubsection{Remaining Numerical Integral}

With the divergent contributions explicitly isolated and calculated, the remaining contribution at $1/\epsilon$ order can be expressed as a finite, numerical integral, just as for the Abelian contribution.  One thing to note about the measurement function of \Eq{eq:wta0eps} is that by setting $\epsilon = 0$, the soft $q\bar q$ contribution is explicitly absent, as noted above.  Performing the integration over the full matrix element and measurement function with $\epsilon = 0$ and all relevant soft and collinear divergences subtracted, we find
\begin{align}
J^{(2,\text{bare,num})}_{g\to q,\text{nab}}(\mu_0^2,k_\perp^2) &=\left(
\frac{\alpha_s}{2\pi}
\right)^2C_A T_R\left(\frac{\mu_0^2}{k_\perp^2}\right)^{2\epsilon}\frac{0.369}{\epsilon}\,,
\end{align}
with error at the order of the last quoted digit.

\subsubsection{Total Real Emission Contribution}

Summing together the soft gluon and $q\bar q$ collinear subtraction terms and the result of the numerical integral, the contribution to the bare jet function is
\begin{align}\label{eq:nab2loopgq}
J^{(2,\text{bare, sub})}_{g\to q,\text{nab}}(\mu_0^2,k_\perp^2) &=\left(
\frac{\alpha_s}{2\pi}
\right)^2\left(\frac{\mu_0^2}{k_\perp^2}\right)^{2\epsilon}\frac{C_A T_R}{\epsilon^2}\\
&\hspace{2cm}\times\left[
-\frac{1}{6\epsilon}-\frac{5}{36}+\frac{2}{3}\log 2+\epsilon\left(
-\frac{107}{216}+\frac{5}{18}\log 2-\frac{\pi^2}{36}+0.369
\right)
\right]\,.\nonumber
\end{align}

\subsection{One-Loop Splitting Function}

In addition to the $1\to 3$ real splitting contribution, we also need to include contributions from the $1\to 2$ one-loop splitting functions that can contribute to the $g\to q$ flavor transition.  The corresponding one-loop splitting function is \cite{Bern:1994zx,Bern:1994fz,Bern:1995ix,Bern:1998sc,Kosower:1999rx}
\begin{align}
P_{g\to q\bar q}^{(1)} &= \left(
\frac{\alpha_s}{2\pi}
\right)^2\left(
\frac{\mu_0^2}{k_\perp^2}
\right)^{2\epsilon} T_R\frac{(4\pi)^2}{s^{1+\epsilon}}\left(
1-2\frac{z(1-z)}{1-\epsilon}
\right)\left[C_F\left(
-\frac{2}{\epsilon^2}-\frac{3}{\epsilon}+\frac{4}{3}\pi^2-8+\cdots
\right)\right.\\
&\hspace{1cm}+C_A\left(
\frac{1}{\epsilon^2}+\frac{11}{3}\frac{1}{\epsilon}+\frac{\log z+\log(1-z)}{\epsilon}-\frac{1}{2}\log^2\frac{1-z}{z}+\frac{76}{9}-\frac{3}{4}\pi^2+\cdots
\right)\nonumber\\
&\hspace{1cm}\left.+\,n_f T_R\left(
-\frac{4}{3}\frac{1}{\epsilon}-\frac{20}{9}+\cdots
\right)\right]\nonumber\,.
\end{align}
In this expression, we have rescaled the invariant mass $s\to k_\perp^2s$, so that $s$ is now dimensionless and $z$ is the anti-quark's energy fraction.  Differential two-body phase space in these coordinates is
\begin{align}
d\Phi^{(2)} = ds\, dz\,\frac{z^{-\epsilon}(1-z)^{-\epsilon}s^{-\epsilon}}{(4\pi)^{2-\epsilon}\Gamma(1-\epsilon)}\,,
\end{align}
and the measurement function that ensures that the quark is higher energy and resolved from the anti-quark is
\begin{align}
\Theta_\text{WTA} = \Theta\left(s - \frac{1-z}{z}\right)\Theta\left(\frac{1}{2}-z\right)\,.
\end{align}
Integrating the splitting functions over phase space with the measurement function, we find the contribution to the jet function to be
\begin{align}\label{eq:1loopsplitgq}
J^{(1,\text{bare,1-loop})}_{g\to q,\text{ab}}(\mu_0^2,k_\perp^2) &= \left(
\frac{\alpha_s}{2\pi}
\right)^2T_R\left(\frac{\mu_0^2}{k_\perp^2}\right)^{2\epsilon}\\
&\hspace{-1cm}\times\left[C_F\left(
-\frac{1}{3\epsilon^3}+\frac{1}{\epsilon^2}\left(
-\frac{8}{9}+\frac{4}{3}\log 2
\right)+\frac{1}{\epsilon}\left(
-\frac{487}{108}+\frac{\pi^2}{18}+\frac{41}{9}\log 2+\frac{4}{3}\log^2 2\right)\right)\right.\nonumber\\
&\hspace{0cm}+C_A\left(
\frac{1}{6\epsilon^3}+\frac{1}{\epsilon^2}\left(
\frac{4}{9}-\frac{2}{3}\log 2
\right)+\frac{1}{\epsilon}\left(
\frac{79}{54}-\frac{\pi^2}{36}-\frac{19}{9}\log 2\right)
\right)\nonumber\\
&\hspace{0cm}\left.+\,n_f T_R\left(
-\frac{2}{9}\frac{1}{\epsilon^2} + \frac{1}{\epsilon}\left(
-\frac{17}{27}+\frac{8}{9}\log 2
\right)
\right)
\right]\nonumber\,.
\end{align}

\subsection{Complete Bare Jet Function at Two-Loop Order}

We now collect the results to establish the complete bare two-loop jet function for the flavor transition $g\to q$.  Summing the results of Eqs.~\ref{eq:ab2loopgq}, \ref{eq:nab2loopgq}, and \ref{eq:1loopsplitgq}, we have
\begin{align}
J^{(2,\text{bare})}_{g\to q}(\mu_0^2,k_\perp^2) &= \left(
\frac{\alpha_s}{2\pi}
\right)^2\left(\frac{\mu_0^2}{k_\perp^2}\right)^{2\epsilon}\frac{T_R}{\epsilon^2}\\
&\hspace{1cm}\times\left[
C_F\left(
\frac{5}{48} - \frac{1}{3}\log 2 
+\epsilon\left(
\frac{7}{576}+\frac{\pi^2}{18}-\frac{67}{72}\log 2-\frac{2}{3}\log^2 2-0.6378
\right)\right)\right.\nonumber\\
&\hspace{2cm}+C_A\left(
\frac{11}{36}+\epsilon\left(
\frac{209}{216}+\frac{\pi^2}{36}-\frac{11}{6}\log 2+0.369
\right)
\right)\nonumber\\
&\hspace{2cm}\left.+\,n_fT_R\left(
-\frac{2}{9} + \epsilon\left(
-\frac{17}{27}+\frac{8}{9}\log 2
\right)
\right)
\right]\nonumber\,.
\end{align}
Then, using the renormalization prescription as described in \Sec{sec:review}, anomalous dimensions and renormalization factors can be extracted from this bare jet function.  Extremely importantly for the validity of the factorization theorem, all triple poles, $1/\epsilon^3$, have explicitly canceled and the remaining double poles arise exclusively due to iterated, strongly-ordered $1\to2$ collinear splittings.  All necessary ingredients for renormalization at one-loop order that have been calculated elsewhere are presented in the appendices.

\section{Bare Jet Function Calculations for Other Flavor Transitions}\label{sec:barejet}

For the remaining bare jet function calculations we will present fewer details, just reporting measurement functions, splitting functions, and results.  With the sum rules and symmetries of QCD, there are only three more unique flavor transitions that must be calculated: $q\to g$, $q\to q'$ with $q'\neq q,\bar q$, and $q\to \bar q$.

\subsection{$\gamma_{q\to g}$}

\subsubsection{$q\to \bar q'q'q$}

For the $q\to g$ transition, there are two distinct final states that can contribute: $q\to \bar q' q' q$ and $q\to ggq$.  Starting with the three-quark final state, the measurement function is especially simple because the only way that the WTA flavor can be a gluon is if the $q'\bar q'$ pair are unresolved:
\begin{align}
\Theta_{\text{WTA},\bar q'_1q'_2q_3} &= \Theta(\min[z_1^2,z_3^2]\theta_{13}^2 - \min[z_1^2,z_2^2]\theta_{12}^2)\Theta(\min[z_2^2,z_3^2]\theta_{23}^2 - \min[z_1^2,z_2^2]\theta_{12}^2)\\
&\hspace{1cm}\times\Theta\left(\frac{1}{2}-z_3\right)\Theta(k_\perp^2 - \min[z_1^2,z_2^2]\theta_{12}^2 E^2)\nonumber\\
&\hspace{-1cm}\times\left[
\Theta(z_1-z_2)\Theta(z_3^2\theta_{13}^2E^2-k_\perp^2)\Theta(z_3^2\theta_{13}^2-z_2^2\theta_{12}^2)+\Theta(z_2-z_1)\Theta(z_3^2\theta_{23}^2E^2-k_\perp^2)\Theta(z_3^2\theta_{23}^2-z_1^2\theta_{12}^2)
\right]\nonumber\,.
\end{align}
The only matrix element that contributes to the divergences of the bare jet function treats all possible quarks $q'$ as distinguishable from $q$:
\begin{align}
|{\cal M}_{q\to \bar q'_1q'_2 q_3}|^2 &= \left(
\frac{\alpha_s}{2\pi}
\right)^2\frac{C_F n_fT_R}{2}\frac{(4\pi)^4(\mu_0^2)^{2\epsilon}}{z_1z_2\theta_{12}^2\left(z_1z_2\theta_{12}^2+z_1z_3\theta_{13}^2+z_2z_3\theta_{23}^2\right)}\\
&\hspace{0.5cm}\times\Bigg[
-\frac{z_1z_2\theta^2_{12}}{z_1z_2\theta_{12}^2+z_1z_3\theta_{13}^2+z_2z_3\theta_{23}^2}\left(
\frac{2z_3}{1-z_3}\frac{\theta_{23}^2-\theta_{13}^2}{\theta_{12}^2}+\frac{z_1-z_2}{z_1+z_2}
\right)^2\nonumber\\
&\hspace{1cm}+\frac{4z_3+(z_1-z_2)^2}{z_1+z_2}+(1-2\epsilon)\left(
z_1+z_2-\frac{z_1z_2\theta_{12}^2}{z_1z_2\theta_{12}^2+z_1z_3\theta_{13}^2+z_2z_3\theta_{23}^2}
\right)
\Bigg]\,.\nonumber
\end{align}
Upon integrating over the overall angular scale of the splitting and rescaling by $\theta_{13}^2$, the measurement function becomes
\begin{align}
\Theta_{\text{WTA},\bar q'_1q'_2q_3} &= \Theta(\min[z_1^2,z_3^2] - \min[z_1^2,z_2^2]\theta_{12}^2)\Theta(\min[z_2^2,z_3^2]\theta_{23}^2 - \min[z_1^2,z_2^2]\theta_{12}^2)\Theta\left(\frac{1}{2}-z_3\right)\\
&\hspace{-1cm}\times\left[
(z_3^{4\epsilon} - (z_2^2\theta_{12}^2)^{2\epsilon})\Theta(z_1-z_2)\Theta(z_3^2-z_2^2\theta_{12}^2)+((z_3^2\theta_{23}^2)^{2\epsilon}-(z_1^2\theta_{12}^2)^{2\epsilon})\Theta(z_2-z_1)\Theta(z_3^2\theta_{23}^2-z_1^2\theta_{12}^2)
\right]\nonumber\,.
\end{align}

Because of the restriction that the $\bar q'q'$ pair must be unresolved, but the $q$ must be resolved from them, only the collinear limit of the matrix element contributes to the divergences of the bare jet function.  The collinear limit $\theta_{12}^2\to 0$ of the measurement function is
\begin{align}
\Theta_{\text{WTA},\bar q'_1q'_2q_3}^\text{coll} &\to \Theta(1-\theta_{12}^2)\Theta\left(\frac{1}{2}-z_3\right)\\
&\hspace{1cm}\times\left[
(z_3^{4\epsilon} - (z_2^2\theta_{12}^2)^{2\epsilon})\Theta(z_1-z_2)+(z_3^{4\epsilon}-(z_1^2\theta_{12}^2)^{2\epsilon})\Theta(z_2-z_1)
\right]\nonumber\,,
\end{align}
while the collinear limit of the matrix element is
\begin{align}
|{\cal M}_{q\to \bar q'_1q'_2 q_3}^\text{coll}|^2 &\to \frac{C_F n_fT_R}{2}\frac{1}{z_1z_2z_3(1-z_3)\theta_{12}^2}\\
&\hspace{0.5cm}\times\Bigg[
-\frac{16z_1z_2z_3}{(1-z_3)^3}
\cos^2\phi+\frac{4z_3+(z_1-z_2)^2}{z_1+z_2}+(1-2\epsilon)
(z_1+z_2)
\Bigg]\,.\nonumber
\end{align}
This collinear limit can then be integrated over three-body phase space.

\subsubsection{$q\to ggq$}

Next, we consider the contribution of the splitting $q\to g_1g_2q_3$ to the flavor transition $q\to g$.  The measurement function is much more complicated now, and all possible clusterings must be considered:
\begin{align}
\Theta_{\text{WTA},g_1g_2q_3} &= \Theta(\theta_{13}^2-\theta_{23}^2)\left\{\Theta(\min[z_1^2,z_3^2]\theta_{13}^2 - \min[z_1^2,z_2^2]\theta_{12}^2)\Theta(\min[z_2^2,z_3^2]\theta_{23}^2 - \min[z_1^2,z_2^2]\theta_{12}^2)\right.
\nonumber\\
&\hspace{1cm}\times \Theta\left(
\frac{1}{2}-z_3
\right)\left[
\Theta(z_1-z_2)\Theta(z_3^2\theta_{13}^2E^2 - k_\perp^2)+\Theta(z_2-z_1)\Theta(z_3^2\theta_{23}^2E^2 - k_\perp^2)
\right]\nonumber\\
&+\Theta(\min[z_1^2,z_2^2]\theta_{12}^2 - \min[z_1^2,z_3^2]\theta_{13}^2)\Theta(\min[z_2^2,z_3^2]\theta_{23}^2 - \min[z_1^2,z_3^2]\theta_{13}^2)\nonumber\\
&\hspace{1cm}\times\left[
\Theta\left(z_2-\frac{1}{2}\right)\left[
\Theta(z_1-z_3)\Theta((1-z_2)^2\theta_{12}^2E^2 - k_\perp^2)+\Theta(z_3-z_1)\Theta((1-z_2)^2\theta_{23}^2E^2 - k_\perp^2)
\right]\right.\nonumber\\
&\hspace{2cm}\left.+\Theta\left(\frac{1}{2}-z_2\right)\Theta(z_1-z_3)\Theta(z_3^2\theta_{13}^2E^2 - k_\perp^2)
\right]\nonumber\\
&+\Theta(\min[z_1^2,z_2^2]\theta_{12}^2 - \min[z_2^2,z_3^2]\theta_{23}^2)\Theta(\min[z_1^2,z_3^2]\theta_{13}^2 - \min[z_2^2,z_3^2]\theta_{23}^2)\nonumber\\
&\hspace{1cm}\times\left[
\Theta\left(z_1-\frac{1}{2}\right)\left[
\Theta(z_2-z_3)\Theta((1-z_1)^2\theta_{12}^2E^2 - k_\perp^2)+\Theta(z_3-z_2)\Theta((1-z_1)^2\theta_{13}^2E^2 - k_\perp^2)
\right]\right.\nonumber\\
&\hspace{2cm}\left.\left.+\Theta\left(\frac{1}{2}-z_1\right)\Theta(z_2-z_3)\Theta(z_3^2\theta_{23}^2E^2 - k_\perp^2)
\right]\right\}
\end{align}
Because gluons $g_1,g_2$ are identical, we have to have a kinematic way to distinguish them.  We do this by stating that $g_1$ is a larger angle away from $q_3$ than $g_2$ from $q_3$.  The matrix elements that contribute to this final state can emit gluons in a correlated (non-Abelian) or uncorrelated (Abelian) manner, where
\begin{align}
|{\cal M}_{q\to g_1g_2 q_3}|^2 = \left(
\frac{\alpha_s}{2\pi}
\right)^2\frac{(4\pi)^4(\mu_0^2)^{2\epsilon}}{(z_1z_2\theta_{12}^2+z_1z_3\theta_{13}^2+z_2z_3\theta_{23}^2)^2}\left[
C_F^2 P^\text{ab}_{q\to g_1g_2 q_3}+C_FC_A P^\text{nab}_{q\to  g_1g_2  q_3}
\right]\,,
\end{align}
with the Abelian splitting function
\begin{align}
P^\text{ab}_{q\to g_1g_2 q_3} &= \frac{(z_1z_2\theta_{12}^2+z_1z_3\theta_{13}^2+z_2z_3\theta_{23}^2)^2}{z_1z_2z_3\theta_{13}^2\theta_{23}^2}\left(
\frac{1+z_3^2}{z_1z_2}-\epsilon\frac{z_1^2+z_2^2}{z_1z_2}-\epsilon(1+\epsilon)
\right)\\
&\hspace{-1cm}+\frac{z_1z_2\theta_{12}^2+z_1z_3\theta_{13}^2+z_2z_3\theta_{23}^2}{z_1z_3\theta_{13}^2}\left(
\frac{z_3(1-z_1)+(1-z_2)^3}{z_1z_2}+\epsilon^2(1+z_3)-\epsilon(z_1^2+z_1z_2+z_2^2)\frac{1-z_2}{z_1z_2}
\right)\nonumber\\
&\hspace{-1cm}+\frac{z_1z_2\theta_{12}^2+z_1z_3\theta_{13}^2+z_2z_3\theta_{23}^2}{z_2z_3\theta_{23}^2}\left(
\frac{z_3(1-z_2)+(1-z_1)^3}{z_1z_2}+\epsilon^2(1+z_3)-\epsilon(z_1^2+z_1z_2+z_2^2)\frac{1-z_1}{z_1z_2}
\right)\nonumber\\
&\hspace{1cm}+(1-\epsilon)\left(
2\epsilon - (1-\epsilon)\frac{z_2\theta_{23}^2}{z_1\theta_{13}^2}- (1-\epsilon)\frac{z_1\theta_{13}^2}{z_2\theta_{23}^2}
\right)
\nonumber\,,
\end{align}
and the non-Abelian splitting function
\begin{align}
P^\text{nab}_{q\to g_1g_2 q_3} &=(1-\epsilon)\left(\frac{1}{2}-\epsilon\right)+\frac{1-\epsilon}{2(\theta_{12}^2)^2}\left(
2z_3\frac{\theta_{23}^2-\theta_{13}^2}{z_1+z_2} + \frac{z_1-z_2}{z_1+z_2}\theta_{12}^2
\right)^2\\
&\hspace{-1cm}+\frac{(z_1z_2\theta_{12}^2+z_1z_3\theta_{13}^2+z_2z_3\theta_{23}^2)^2}{2z_1^2z_2z_3\theta_{12}^2\theta_{13}^2}\left(
\frac{(1-z_3)^2(1-\epsilon)+2z_3}{z_2}+\frac{z_2^2(1-\epsilon)+2(1-z_2)}{1-z_3}
\right)\nonumber\\
&\hspace{-1cm}+\frac{(z_1z_2\theta_{12}^2+z_1z_3\theta_{13}^2+z_2z_3\theta_{23}^2)^2}{2z_1z_2^2z_3\theta_{12}^2\theta_{23}^2}\left(
\frac{(1-z_3)^2(1-\epsilon)+2z_3}{z_1}+\frac{z_1^2(1-\epsilon)+2(1-z_1)}{1-z_3}
\right)\nonumber\\
&\hspace{-1cm}-\frac{(z_1z_2\theta_{12}^2+z_1z_3\theta_{13}^2+z_2z_3\theta_{23}^2)^2}{2z_1z_2z_3\theta_{13}^2\theta_{23}^2}\left(
\frac{(1-z_3)^2(1-\epsilon)+2z_3}{z_1z_2}+\epsilon(1-\epsilon)
\right)\nonumber\\
&\hspace{-1cm}+\frac{z_1z_2\theta_{12}^2+z_1z_3\theta_{13}^2+z_2z_3\theta_{23}^2}{2z_1z_2\theta_{12}^2}\left(
(1-\epsilon)\frac{z_1(2-2z_1+z_1^2)-z_2(6-6z_2+z_2^2)}{z_2(1-z_3)}+2\epsilon\frac{z_3(z_1-2z_2)-z_2}{z_2(1-z_3)}
\right)\nonumber\\
&\hspace{-1cm}+\frac{z_1z_2\theta_{12}^2+z_1z_3\theta_{13}^2+z_2z_3\theta_{23}^2}{2z_1z_2\theta_{12}^2}\left(
(1-\epsilon)\frac{z_2(2-2z_2+z_2^2)-z_1(6-6z_1+z_1^2)}{z_1(1-z_3)}+2\epsilon\frac{z_3(z_2-2z_1)-z_1}{z_1(1-z_3)}
\right)\nonumber\\
&\hspace{-1cm}+\frac{z_1z_2\theta_{12}^2+z_1z_3\theta_{13}^2+z_2z_3\theta_{23}^2}{2z_1z_3\theta_{13}^2}\left[
(1-\epsilon)\frac{(1-z_2)^3+z_3^2-z_2}{z_2(1-z_3)}-\epsilon\left(
\frac{2(1-z_2)(z_2-z_3)}{z_2(1-z_3)}-z_1+z_2
\right)\right.\nonumber\\
&\hspace{5cm}\left.-\frac{z_3(1-z_1)+(1-z_2)^3}{z_1z_2}+\epsilon(1-z_2)\left(
\frac{z_1^2+z_2^2}{z_1z_2}-\epsilon
\right)
\right]\nonumber\\
&\hspace{-1cm}+\frac{z_1z_2\theta_{12}^2+z_1z_3\theta_{13}^2+z_2z_3\theta_{23}^2}{2z_2z_3\theta_{23}^2}\left[
(1-\epsilon)\frac{(1-z_1)^3+z_3^2-z_1}{z_1(1-z_3)}-\epsilon\left(
\frac{2(1-z_1)(z_1-z_3)}{z_1(1-z_3)}-z_2+z_1
\right)\right.\nonumber\\
&\hspace{5cm}\left.-\frac{z_3(1-z_2)+(1-z_1)^3}{z_1z_2}+\epsilon(1-z_1)\left(
\frac{z_1^2+z_2^2}{z_1z_2}-\epsilon
\right)
\right]\nonumber\,.
\end{align}
Rescaling and integrating over the overall angle of the splitting $\theta_{13}^2$, the measurement function becomes
\begin{align}
\Theta_{\text{WTA},g_1g_2q_3} &= \Theta(1-\theta_{23}^2)\left\{\Theta(\min[z_1^2,z_3^2] - \min[z_1^2,z_2^2]\theta_{12}^2)\Theta(\min[z_2^2,z_3^2]\theta_{23}^2 - \min[z_1^2,z_2^2]\theta_{12}^2)\right.
\nonumber\\
&\hspace{1cm}\times \Theta\left(
\frac{1}{2}-z_3
\right)\left[
z_3^{4\epsilon}\Theta(z_1-z_2)+(z_3^2\theta_{23}^2)^{2\epsilon}\Theta(z_2-z_1)
\right]\nonumber\\
&+\Theta(\min[z_1^2,z_2^2]\theta_{12}^2 - \min[z_1^2,z_3^2])\Theta(\min[z_2^2,z_3^2]\theta_{23}^2 - \min[z_1^2,z_3^2])\nonumber\\
&\hspace{1cm}\times\left[
\Theta\left(z_2-\frac{1}{2}\right)\left[
((1-z_2)^2\theta_{12}^2)^{2\epsilon}\Theta(z_1-z_3)+((1-z_2)^2\theta_{23}^2)^{2\epsilon}\Theta(z_3-z_1)
\right]\right.\nonumber\\
&\hspace{2cm}\left.+z_3^{4\epsilon}\Theta\left(\frac{1}{2}-z_2\right)\Theta(z_1-z_3)
\right]\nonumber\\
&+\Theta(\min[z_1^2,z_2^2]\theta_{12}^2 - \min[z_2^2,z_3^2]\theta_{23}^2)\Theta(\min[z_1^2,z_3^2] - \min[z_2^2,z_3^2]\theta_{23}^2)\nonumber\\
&\hspace{1cm}\times\left[
\Theta\left(z_1-\frac{1}{2}\right)\left[
((1-z_1)^2\theta_{12}^2)^{2\epsilon}\Theta(z_2-z_3)+(1-z_1)^{4\epsilon}\Theta(z_3-z_2)
\right]\right.\nonumber\\
&\hspace{2cm}\left.\left.+(z_3^2\theta_{23}^2)^{2\epsilon}\Theta\left(\frac{1}{2}-z_1\right)\Theta(z_2-z_3)
\right]\right\}\,.
\end{align}

With these expressions, there are ostensibly 5 singular limits to consider: either gluon going soft, $z_1,z_2\to 0$, gluons becoming collinear to each other, $\theta_{12}^2$, or the gluons becoming collinear to the quark $\theta_{13}^2,\theta_{23}^2\to 0$.  We have explicitly eliminated the possibility of $\theta_{13}^2\to 0$ in the way we have broken indistinguishability of the gluons, so we need not consider this limit.  For the soft limits, note that the measurement function reduces to
\begin{align}
\Theta_{\text{WTA},g_1g_2q_3}^{z_1\to 0} &\to (z_3^2\theta_{23}^2)^{2\epsilon}\Theta(1-\theta_{23}^2)\Theta\left(
\frac{1}{2}-z_3
\right)\,,\\
\Theta_{\text{WTA},g_1g_2q_3}^{z_2\to 0}&\to z_3^{4\epsilon}\Theta(1-\theta_{23}^2)
\Theta\left(
\frac{1}{2}-z_3
\right)\,.\nonumber
\end{align}
The only difference between these measurement functions is the power of $\theta_{23}^2$, and because the gluons are identical the corresponding soft matrix elements are identical, up to the interchange $1\leftrightarrow 2$.  We can, then, change variables in the $z_1\to 0$ limit with $\theta_{23}^2 \to 1/\theta_{23}^2$, and then the measurement functions become
\begin{align}
\Theta_{\text{WTA},g_1g_2q_3}^{z_1\to 0} &\to z_3^{4\epsilon}\Theta(\theta_{23}^2-1)\Theta\left(
\frac{1}{2}-z_3
\right)\,,\\
\Theta_{\text{WTA},g_1g_2q_3}^{z_2\to 0}&\to z_3^{4\epsilon}\Theta(1-\theta_{23}^2)
\Theta\left(
\frac{1}{2}-z_3
\right)\,.\nonumber
\end{align}
The sum of these two measurement functions is then completely inclusive over the angle $\theta_{23}^2$, and therefore the Abelian contribution will explicitly vanish because there is a scaleless integral over $\theta_{23}^2$.  The soft limit of the non-Abelian matrix element will have a non-zero contribution because it lacks a collinear divergence $\theta_{23}^2\to 0$.  The soft limit of the non-Abelian matrix element is
\begin{align}
|{\cal M}_{q\to g_1g_2 q_3,\text{nab}}^{z_1\to 0}|^2 \to \frac{C_FC_A}{z_1^2z_3(1-z_3)}\frac{1+\theta_{23}^2-\theta_{12}^2}{\theta_{12}^2\theta_{23}^2}\left(
\frac{1+z_3^2}{1-z_3}-\epsilon(1-z_3)
\right)\,,
\end{align}
or $1\leftrightarrow 2$.

For the collinear limits, we first consider $\theta_{23}^2\to 0$.  In this limit the measurement function becomes
\begin{align}
\Theta_{\text{WTA},g_1g_2q_3}^\text{ab,coll} &\to \Theta(1-\theta_{23}^2)\left[
(1-z_1)^{4\epsilon}\Theta\left(z_1-\frac{1}{2}\right)+(z_3^2\theta_{23}^2)^{2\epsilon}\Theta\left(\frac{1}{2}-z_1\right)\Theta(z_2-z_3)
\right]\,.
\end{align}
In this collinear limit, only the Abelian matrix element has a divergence, where
\begin{align}
|{\cal M}_{q\to g_1g_2q_3}^\text{ab,coll}|^2 &\to \frac{C_F^2}{z_1z_2z_3(1-z_1)^2\theta_{23}^2}\left[(1-z_1)^2\left(
\frac{1+z_3^2}{z_1z_2}-\epsilon\frac{z_1^2+z_2^2}{z_1z_2}-\epsilon(1+\epsilon)
\right)\right.\\
&\hspace{-1cm}\left.+(1-z_1)\left(
\frac{z_3(1-z_2)+(1-z_1)^3}{z_1z_2}+\epsilon^2(1+z_3)-\epsilon(z_1^2+z_1z_2+z_2^2)\frac{1-z_1}{z_1z_2}
\right)-(1-\epsilon)^2
z_3\right]\,.
\nonumber
\end{align}
For the collinear limit $\theta_{12}^2$, only the non-Abelian matrix element has a divergence.  The measurement function reduces to\begin{align}
\Theta_{\text{WTA},g_1g_2q_3}^\text{nab,coll} &\to z_3^{4\epsilon}\Theta(\cos\phi)\Theta(1-\theta_{12}^2) \Theta\left(
\frac{1}{2}-z_3
\right)\,,
\end{align}
and the matrix element becomes
\begin{align}
|{\cal M}_{q\to g_1g_2q_3}^\text{nab,coll}|^2 &\to \frac{C_FC_A}{z_3^2(1-z_3)^2\theta_{12}^2}\left[(1-\epsilon)\frac{8z_3^2}{(1-z_3)^2}
\cos^2\phi\right.\\
&\hspace{-1cm}+\frac{z_3(1-z_3)^2}{2z_1^2z_2}\left(
\frac{(1-z_3)^2(1-\epsilon)+2z_3}{z_2}+\frac{z_2^2(1-\epsilon)+2(1-z_2)}{1-z_3}
\right)\nonumber\\
&\hspace{-1cm}+\frac{z_3(1-z_3)^2}{2z_1z_2^2}\left(
\frac{(1-z_3)^2(1-\epsilon)+2z_3}{z_1}+\frac{z_1^2(1-\epsilon)+2(1-z_1)}{1-z_3}
\right)\nonumber\\
&\hspace{-1cm}+\frac{z_3(1-z_3)}{2z_1z_2}\left(
(1-\epsilon)\frac{z_1(2-2z_1+z_1^2)-z_2(6-6z_2+z_2^2)}{z_2(1-z_3)}+2\epsilon\frac{z_3(z_1-2z_2)-z_2}{z_2(1-z_3)}
\right)\nonumber\\
&\hspace{-1cm}\left.+\frac{z_3(1-z_3)}{2z_1z_2}\left(
(1-\epsilon)\frac{z_2(2-2z_2+z_2^2)-z_1(6-6z_1+z_1^2)}{z_1(1-z_3)}+2\epsilon\frac{z_3(z_2-2z_1)-z_1}{z_1(1-z_3)}
\right)\nonumber\right]
\nonumber\,.
\end{align}
Integrating these matrix elements and measurement functions on phase space produce the singular real contributions to the bare jet function.  To calculate the non-singular contributions to the bare jet function, we perform the same procedure as for the transition $g\to q$ earlier.  We take the $\epsilon = 0$ limit of phase space, measurement functions and matrix elements, and perform the numerical integral over the full measurement function and matrix element, with all singular limits explicitly subtracted.

\subsubsection{One-Loop Splitting Function $q\to gq$}

We also must consider the contributions to the bare jet function from the one-loop splitting function $q\to gq$.  The one-loop splitting function is
\begin{align}
P_{q\to qg}^{(1)} &= \left(
\frac{\alpha_s}{2\pi}
\right)^2\left(
\frac{\mu_0^2}{k_\perp^2}
\right)^{2\epsilon} C_F\frac{(4\pi)^2}{s^{1+\epsilon}}\left(
\frac{1+(1-z)^2}{z}-\epsilon z
\right)\\
&\hspace{-1cm}\times\left[C_F\left(
\frac{2}{\epsilon}\log(1-z)-\log^2(1-z)-2\,\text{Li}_2(z)+\frac{z}{1+(1-z)^2}
\right)\right.\nonumber\\
&\hspace{-1cm}\left.+C_A\left(
-\frac{1}{\epsilon^2}+\frac{1}{\epsilon}\log\frac{z}{1-z}-\frac{\pi^2}{12}+\frac{1}{2}\log^2(1-z)+\log z\log(1-z)-\frac{1}{2}\log^2 z+2\,\text{Li}_2(z)+\frac{z}{1+(1-z)^2}
\right)\right]\nonumber\,,
\end{align}
where here, $z$ is the energy fraction of the final state gluon and, as earlier, we have rescaled the invariant mass $s\to k_\perp^2 s$.  The measurement function that enforces that the gluon is hard and resolved is then
\begin{align}
\Theta_{\text{WTA},gq} = \Theta\left(s - \frac{z}{1-z}\right)\Theta\left(z-\frac{1}{2}\right)\,.
\end{align}
This can then be integrated over two-body phase space to determine its contribution to the bare jet function.

\subsubsection{Complete Bare Jet Function}

Putting all of these pieces together, we find the total two-loop bare jet function for the flavor transition $q\to g$ to be
\begin{align}
J^{(2,\text{bare})}_{q\to g} &= \left(
\frac{\alpha_s}{2\pi}
\right)^2\left(
\frac{\mu_0^2}{k_\perp^2}
\right)^{2\epsilon}\frac{C_F}{\epsilon^2}\\
&\hspace{-2cm}\times\left[
C_F\left(
-\frac{1}{2}\left(
\frac{5}{8}-2\log 2
\right)^2+\epsilon\left(
-\frac{139}{128}+\frac{371}{64}\log 2+\frac{13}{32}\pi^2-2\log^22+\frac{\pi^2}{6}\log 2-\frac{8}{3}\log^3 2-\frac{9}{4}\zeta_3+0.585
\right)
\right)\right.\nonumber\\
&+C_A\left(
\frac{11}{12}\left(
-\frac{5}{8}+2\log 2
\right)+\epsilon\left(
-\frac{593}{144}-\frac{2}{9}\pi^2+\frac{195}{16}\log 2-\frac{5}{6}\pi^2\log 2+\frac{13}{12}\log^2 2+2\zeta_3-0.749
\right)
\right)\nonumber\\
&\left.+\,n_f T_R\left(
\frac{5}{12}-\frac{4}{3}\log 2+\epsilon\left(
\frac{277}{144}-\frac{35}{9}\log 2+\frac{4}{3}\log^22
\right)
\right)
\right]\,.\nonumber
\end{align}
Note that all soft divergences have explicitly canceled and the structure of the double poles in $\epsilon$ is exactly that of iterated strongly-ordered $1\to 2$ collinear splittings.  This bare jet function can then be renormalized as discussed above to determine anomalous dimensions and renormalization factors.

\subsection{$\gamma_{q\to q'}$}

Starting at two-loop order, there are two quark flavor transitions that are no longer forbidden and so their anomalous dimensions start at order-$\alpha_s^2$.  We first consider the transition $q\to q'$, where $q'\neq q,\bar q$, for which there is a single $1\to 3$ splitting that can contribute, $q\to \bar q_1'q_2'q_3$.  At least at a heuristic level, this transition was discussed in Ref.~\cite{Larkoski:2023upz} as justification for the factorization theorem.  There, the measurement function was established to be
\begin{align}
\Theta_\text{WTA}&=\Theta\left(\min[z_1^2,z_3^2]\theta_{13}^2-z_1^2\theta_{12}^2\right)\Theta\left(\min[z_2^2,z_3^2]\theta_{23}^2-z_1^2\theta_{12}^2\right)\\
&\hspace{2cm}\times\Theta\left(\frac{1}{2}-z_3\right)\Theta(z_2-z_1)\Theta\left(z_1^2\theta_{12}^2 E^2-k_\perp^2\right)\nonumber\\
&
\hspace{1cm}+\Theta\left(\min[z_1^2,z_2^2]\theta_{12}^2-z_3^2\theta_{23}^2\right)\Theta\left(\min[z_1^2,z_3^2]\theta_{13}^2-z_3^2\theta_{23}^2\right)\nonumber\\
&\hspace{2cm}\times\Theta\left(\frac{1}{2}-z_1\right)\Theta(z_2-z_3)\Theta\left(z_3^2\theta_{23}^2 E^2-k_\perp^2\right)\nonumber\\
&
\hspace{1cm}+\Theta\left(z_1^2\theta_{12}^2-\min[z_1^2,z_3^2]\theta_{13}^2\right)\Theta\left(z_3^2\theta_{23}^2-\min[z_1^2,z_3^2]\theta_{13}^2\right)\Theta\left(z_2-\frac{1}{2}\right)\nonumber\\
&
\hspace{2cm}
\times\left[
\Theta(z_1-z_3)\Theta\left((1-z_2)^2\theta_{12}^2E^2-k_\perp^2\right)+\Theta(z_3-z_1)\Theta\left((1-z_2)^2\theta_{23}^2 E^2-k_\perp^2\right)
\right]\,.
\nonumber
\end{align}
The matrix element for this splitting is
\begin{align}\label{eq:matelqqq}
|{\cal M}_{q\to \bar q'_1q'_2 q_3}|^2 &= \left(
\frac{\alpha_s}{2\pi}
\right)^2\frac{C_F T_R}{2}\frac{(4\pi)^4(\mu_0^2)^{2\epsilon}}{z_1z_2\theta_{12}^2\left(z_1z_2\theta_{12}^2+z_1z_3\theta_{13}^2+z_2z_3\theta_{23}^2\right)}\\
&\hspace{0.5cm}\times\Bigg[
-\frac{z_1z_2\theta^2_{12}}{z_1z_2\theta_{12}^2+z_1z_3\theta_{13}^2+z_2z_3\theta_{23}^2}\left(
\frac{2z_3}{1-z_3}\frac{\theta_{23}^2-\theta_{13}^2}{\theta_{12}^2}+\frac{z_1-z_2}{z_1+z_2}
\right)^2\nonumber\\
&\hspace{1cm}+\frac{4z_3+(z_1-z_2)^2}{z_1+z_2}+(1-2\epsilon)\left(
z_1+z_2-\frac{z_1z_2\theta_{12}^2}{z_1z_2\theta_{12}^2+z_1z_3\theta_{13}^2+z_2z_3\theta_{23}^2}
\right)
\Bigg]\,.\nonumber
\end{align}
If we rescale all angles by $\theta_{23}^2$ and then integrate over the overall angular scale, the measurement function becomes
with measurement function
\begin{align}
\Theta_\text{WTA}&=(z_1^2\theta_{12}^2)^{2\epsilon}\Theta\left(\min[z_1^2,z_3^2]\theta_{13}^2-z_1^2\theta_{12}^2\right)\Theta\left(\min[z_2^2,z_3^2]-z_1^2\theta_{12}^2\right)\Theta\left(\frac{1}{2}-z_3\right)\Theta(z_2-z_1)
\nonumber\\
&
\hspace{1cm}+z_3^{4\epsilon}\Theta\left(\min[z_1^2,z_2^2]\theta_{12}^2-z_3^2\right)\Theta\left(\min[z_1^2,z_3^2]\theta_{13}^2-z_3^2\right)\Theta\left(\frac{1}{2}-z_1\right)\Theta(z_2-z_3)\nonumber\\
&
\hspace{1cm}+(1-z_2)^{4\epsilon}\Theta\left(z_1^2\theta_{12}^2-\min[z_1^2,z_3^2]\theta_{13}^2\right)\Theta\left(z_3^2-\min[z_1^2,z_3^2]\theta_{13}^2\right)\Theta\left(z_2-\frac{1}{2}\right)\nonumber\\
&
\hspace{2cm}
\times\left[
(\theta_{12}^2)^{2\epsilon}\Theta(z_1-z_3)+\Theta(z_3-z_1)
\right]\,.
\end{align}

There is only one singular limit that needs to be addressed, the collinear limit where the final state $\bar q'q'$ pair becomes collinear, $\theta_{12}^2\to 0$.  The measurement function reduces to
\begin{align}
\Theta_\text{WTA}^\text{coll}&\to(z_1^2\theta_{12}^2)^{2\epsilon}\Theta\left(1-\theta_{12}^2\right)\Theta\left(\frac{1}{2}-z_3\right)\Theta(z_2-z_1)\,,
\end{align}
while the matrix element becomes
\begin{align}
|{\cal M}_{q\to \bar q'_1q'_2 q_3}^\text{coll}|^2 &\to \frac{C_FT_R}{2}\frac{1}{z_1z_2z_3(1-z_3)\theta_{12}^2}\\
&\hspace{0.5cm}\times\Bigg[
-\frac{16z_1z_2z_3}{(1-z_3)^3}\cos^2\phi+\frac{4z_3+(z_1-z_2)^2}{z_1+z_2}+(1-2\epsilon)\left(
z_1+z_2
\right)
\Bigg]\,.\nonumber
\end{align}
The collinear contribution to the bare jet function can be calculated by integrating these expressions over phase space.  The remaining contribution to the $1/\epsilon$ divergence can be calculated again as a numerical integral by setting $\epsilon = 0$ everywhere except exactly at the pole, and then explicitly subtracting off the collinear limit.

Then, the total bare jet function for this flavor transition is
\begin{align}
J^{(2,\text{bare})}_{q\to q'}(\mu_0^2,k_\perp^2) 
&=\left(
\frac{\alpha_s}{2\pi}
\right)^2\left(
\frac{\mu_0^2}{k_\perp^2}
\right)^{2\epsilon}\frac{C_F T_R}{\epsilon^2}\\
&\hspace{1cm}\times\left[
-\frac{5}{48}+\frac{1}{3}\log 2+\epsilon\left(
-\frac{7}{48}-\frac{1}{8}\log 2-\log^2 2+\frac{\pi^2}{18}-0.0223
\right)
\right]\nonumber\,.
\end{align}
As validated in Ref.~\cite{Larkoski:2023upz}, the double poles in $\epsilon$ are exactly described by iterated $1\to 2$ strongly-ordered collinear splittings.

\subsection{$\gamma_{q\to \bar q}$}

The final flavor transition to consider is $q\to \bar q$, quark to its anti-particle transition.  The final state that contributes now consists of two identical quarks, $q\to \bar q_1 q_2q_3$ with measurement function
\begin{align}
\Theta_\text{WTA} &= \Theta(\theta_{13}^2 - \theta_{12}^2)\left\{\Theta(\min[z_1^2,z_3^2]\theta_{13}^2 - z_2^2\theta_{12}^2)\Theta(\min[z_2^2,z_3^2]\theta_{23}^2 - z_2^2\theta_{12}^2)\Theta\left(\frac{1}{2}-z_3\right)\right.\\
&\hspace{1cm}\times\Theta(z_1-z_2)\Theta(z_2^2\theta_{12}^2E^2 - k_\perp^2)\nonumber\\
&+\Theta(\min[z_1^2,z_2^2]\theta_{12}^2 - z_3^2\theta_{13}^2)\Theta(\min[z_2^2,z_3^2]\theta_{23}^2 - z_3^2\theta_{13}^2)\Theta\left(\frac{1}{2}-z_2\right)\nonumber\\
&\hspace{1cm}\times\Theta(z_1-z_3)\Theta(z_3^2\theta_{13}^2E^2 - k_\perp^2)\nonumber\\
&+\Theta(z_2^2\theta_{12}^2-\min[z_2^2,z_3^2]\theta_{23}^2)\Theta(z_3^2\theta_{13}^2-\min[z_2^2,z_3^2]\theta_{23}^2)\Theta\left(z_1-\frac{1}{2}\right)\nonumber\\
&\hspace{1cm}\left.\times\left[
\Theta(z_2-z_3)\Theta((1-z_1)^2\theta_{12}^2E^2-k_\perp^2)+\Theta(z_3-z_2)\Theta((1-z_1)^2\theta_{13}^2E^2-k_\perp^2)
\right]\right\}\nonumber
\end{align}
Because quarks $q_2,q_3$ are identical, we have to have a kinematic way to distinguish them.  We do this by stating that $q_3$ is a larger angle away from $q_1$ than $q_2$ from $q_1$, similarly to what we did with the $ggq$ final state.  Because of the identical particles in the final state, the structure of the matrix element is a bit special, with
\begin{align}
|{\cal M}_{q\to \bar q_1q_2 q_3}|^2 = \left(
\frac{\alpha_s}{2\pi}
\right)^2\frac{(4\pi)^4(\mu_0^2)^{2\epsilon}C_F}{(z_1z_2\theta_{12}^2+z_1z_3\theta_{13}^2+z_2z_3\theta_{23}^2)^2}\left[
T_R P^\text{dist}_{q\to \bar q_1q_2 q_3}+\left(
C_F - \frac{C_A}{2}
\right) P^\text{ind}_{q\to \bar q_1q_2 q_3}
\right]\,,
\end{align}
where the two splitting function contributions we have labeled ``distinguishable'' (dist) and ``indistinguishable'' (ind), and have distinct color factors.  The splitting functions are
\begin{align}
P^\text{dist}_{q\to \bar q_1q_2 q_3} &= \frac{z_1z_2\theta_{12}^2+z_1z_3\theta_{13}^2+z_2z_3\theta_{23}^2}{2z_1z_2\theta_{12}^2}\left[
-\frac{z_1z_2z_3^2\theta_{12}^2\left(
2\frac{\theta_{23}^2-\theta_{13}^2}{\theta_{12}^2} + \frac{z_1-z_2}{z_3}
\right)^2}{(1-z_3)^2(z_1z_2\theta_{12}^2+z_1z_3\theta_{13}^2+z_2z_3\theta_{23}^2)}\right.\\
&\hspace{1cm}\left.+\frac{4z_3+(z_1-z_2)^2}{z_1+z_2}+(1-2\epsilon)\left(z_1+z_2-\frac{z_1z_2\theta_{12}^2}{z_1z_2\theta_{12}^2+z_1z_3\theta_{13}^2+z_2z_3\theta_{23}^2}\right)
\right]\nonumber\\
&+\frac{z_1z_2\theta_{12}^2+z_1z_3\theta_{13}^2+z_2z_3\theta_{23}^2}{2z_1z_3\theta_{13}^2}\left[
-\frac{z_1z_3z_2^2\theta_{13}^2\left(
2\frac{\theta_{23}^2-\theta_{12}^2}{\theta_{13}^2} + \frac{z_1-z_3}{z_2}
\right)^2}{(1-z_2)^2(z_1z_2\theta_{12}^2+z_1z_3\theta_{13}^2+z_2z_3\theta_{23}^2)}\right.\nonumber\\
&\hspace{1cm}\left.+\frac{4z_2+(z_1-z_3)^2}{z_1+z_3}+(1-2\epsilon)\left(z_1+z_3-\frac{z_1z_3\theta_{13}^2}{z_1z_2\theta_{12}^2+z_1z_3\theta_{13}^2+z_2z_3\theta_{23}^2}\right)
\right]\nonumber\,,
\end{align}
and
\begin{align}
P^\text{ind}_{q\to \bar q_1q_2 q_3} &= (1-\epsilon)\left(
\frac{2z_3\theta_{23}^2}{z_1\theta_{12}^2}+\frac{2z_2\theta_{23}^2}{z_1\theta_{13}^2} -2 \epsilon
\right)\\
&\hspace{-1cm}+\frac{z_1z_2\theta_{12}^2+z_1z_3\theta_{13}^2+z_2z_3\theta_{23}^2}{z_1z_2\theta_{12}^2}\left[
\frac{1+z_1^2}{1-z_2} - \frac{2z_2}{1-z_3}-\epsilon\left(
\frac{(1-z_3)^2}{1-z_2}+1+z_1-\frac{2z_2}{1-z_3}
\right)-\epsilon^2(1-z_3)
\right]\nonumber\\
&\hspace{-1cm}+\frac{z_1z_2\theta_{12}^2+z_1z_3\theta_{13}^2+z_2z_3\theta_{23}^2}{z_1z_3\theta_{13}^2}\left[
\frac{1+z_1^2}{1-z_3} - \frac{2z_3}{1-z_2}-\epsilon\left(
\frac{(1-z_2)^2}{1-z_3}+1+z_1-\frac{2z_3}{1-z_2}
\right)-\epsilon^2(1-z_2)
\right]\nonumber\\
&\hspace{-1cm}-\frac{(z_1z_2\theta_{12}^2+z_1z_3\theta_{13}^2+z_2z_3\theta_{23}^2)^2}{z_1z_2z_3\theta_{12}^2\theta_{13}^2}\left[
\frac{1+z_1^2}{(1-z_2)(1-z_3)}-\epsilon\left(
1+\frac{1-z_2}{1-z_3}+\frac{1-z_3}{1-z_2}
\right)-\epsilon^2
\right]
\nonumber\,.
\end{align}
Rescaling angles by $\theta_{13}^2$ and then integrating over the overall angular scale, the measurement function becomes
\begin{align}
\Theta_\text{WTA} &= \Theta(1-\theta_{12}^2)\left\{(z_2^2\theta_{12}^2)^{2\epsilon}\Theta(\min[z_1^2,z_3^2] - z_2^2\theta_{12}^2)\Theta(\min[z_2^2,z_3^2]\theta_{23}^2 - z_2^2\theta_{12}^2)\Theta\left(\frac{1}{2}-z_3\right)\Theta(z_1-z_2)\right.\nonumber\\
&+z_3^{4\epsilon}\Theta(\min[z_1^2,z_2^2]\theta_{12}^2 - z_3^2)\Theta(\min[z_2^2,z_3^2]\theta_{23}^2 - z_3^2)\Theta\left(\frac{1}{2}-z_2\right)\Theta(z_1-z_3)\nonumber\\
&+(1-z_1)^{4\epsilon}\Theta(z_2^2\theta_{12}^2-\min[z_2^2,z_3^2]\theta_{23}^2)\Theta(z_3^2-\min[z_2^2,z_3^2]\theta_{23}^2)\Theta\left(z_1-\frac{1}{2}\right)\nonumber\\
&\hspace{1cm}\left.\times\left[
(\theta_{12}^2)^{2\epsilon}\Theta(z_2-z_3)+\Theta(z_3-z_2)
\right]\right\}\,.
\end{align}

Because of the way we have distinguished quarks 2 and 3, there is but a single collinear limit when $\theta_{12}^2\to 0$ for which the measurement function reduces to
\begin{align}
\Theta_\text{WTA}^\text{coll} &\to (z_2^2\theta_{12}^2)^{2\epsilon}\Theta(1-\theta_{12}^2)\Theta\left(\frac{1}{2}-z_3\right)\Theta(z_1-z_2)\,.
\end{align}
In this limit, only the ``distinguishable'' part of the matrix element is singular, with the limit
\begin{align}
|{\cal M}_{q\to \bar q qq}^\text{coll}|^2&\to \frac{C_FT_R}{2z_1z_2z_3(1-z_3)\theta_{12}^2}
\left(
-\frac{16z_1z_2z_3
\cos^2\phi}{(1-z_3)^3}+\frac{4z_3+(z_1-z_2)^2}{z_1+z_2}+(1-2\epsilon)(z_1+z_2)
\right)\,.
\end{align}
As in all other flavor transitions, we integrate this singular limit over three-body phase space, and then determine the remaining contribution to the $1/\epsilon$ pole in the bare jet function through numerical integration by setting $\epsilon = 0$ and explicitly subtracting off the collinear limit.

Then, the two-loop bare jet function for the flavor transition $q\to \bar q$ is
\begin{align}
J_{q\to \bar q}^{(2,\text{bare})}(\mu_0^2,k_\perp^2)&=\left(
\frac{\alpha_s}{2\pi}
\right)^2\left(
\frac{\mu_0^2}{k_\perp^2}
\right)^{2\epsilon}\frac{C_F}{\epsilon^2}\\
&\hspace{-2cm}\times\left[T_R\left(
-\frac{5}{48}+\frac{1}{3}\log 2+\epsilon\left(
-\frac{7}{48}-\frac{1}{8}\log 2-\log^2 2+\frac{\pi^2}{18}-0.0381
\right)
\right)-0.00379\left(
C_F-\frac{C_A}{2}
\right)\right]\nonumber\,.
\end{align}
The double poles in $\epsilon$ correspond to iterated strongly-ordered $1\to 2$ collinear splitting and this expression can be renormalized to extract anomalous dimensions.

\section{Conclusions}\label{sec:concs}

Defining perturbative jet flavor through fragmentation, assuring soft safety but forgoing collinear safety, enables a factorization theorem of hard and jet functions that have especially simple collinear renormalization group evolution.  The Winner-Take-All axis and the corresponding flavor of the particle on which it lies, enables an algorithmically-simple definition of jet flavor that can be implemented into parton shower simulations or high-precision fixed-order results.  In this paper, we have extended previous results, extracting the two-loop anomalous dimensions of the factorization theorem for all possible flavor transitions and through doing so determine all objects necessary for renormalization of hard functions calculated at next-to-next-to-leading order.

In addition to the calculation of the hard function and the extraction of its low-scale matrix elements in all flavor channels, the remaining ingredient for complete next-to-next-to-leading order and next-to-leading logarithmic resummation of WTA jet flavor are the two-loop low-scale matrix elements of the jet function.  With the results established here, their calculation is straightforward, if significant bookkeeping.  Further, even in the singularities of the two-loop bare jet function, we have expressed contributions from non-singular components of matrix elements as the result of a numerical integral, because the phase space constraints are, at least naively, very complicated.  However, for this definition of jet flavor, the phase space measurement constraints are much more inclusive than many other observables calculated to this order, and so it may be possible to completely analytically calculate the bare jet function.  Such an analytic result would have a structure of transcendental numbers that may be familiar from other inclusive calculations, like form factors or fragmentation in ${\cal N} = 4$ Super-Yang Mills, e.g., \cite{Kotikov:2002ab,Kotikov:2004er,Bern:2006ew,Gehrmann:2011xn}, or from the collinear factorization theorem of the energy-energy correlator \cite{Dixon:2019uzg,Chen:2019bpb,Chicherin:2024ifn}.  We look forward to continuing to push the precision boundary of jet flavor with fragmentation.

\section*{Acknowledgements}

I thank the many, many excellent physicists I have worked with in my career for their insight, guidance, patience, and mentorship.

\appendix

\section{Summary of One-Loop Results}

In this appendix, we collect the necessary one-loop results for explicit renormalization of the two-loop bare jet function.  First, the full expanded renormalization equation relating the two-loop bare jet function to its renormalized function is
\begin{align}
J_{j\to f}^{(2,\text{bare})} = J_{j\to f}^{(2,\text{ren})} + \sum_kZ_{j\to k}^{(1)}J_{k\to f}^{(1,\text{ren})} + Z_{j\to k}^{(2)} + \frac{\alpha_s}{2\pi}\beta_0 \frac{1}{\epsilon}\left(\frac{\mu_0^2}{\mu^2}\right)^\epsilon J_{j\to f}^{(1,\text{ren})}\,.
\end{align}
Here, the numerical superscript denotes the loop order of the expression.  

From explicit solution of the renormalization group equations, the two-loop renormalized jet function takes the form
\begin{align}
&J_{j\to f}(\mu^2,k_\perp^2)\\
&\supset \left(
\frac{\alpha_s}{2\pi}
\right)^2\left[
\frac{\sum_k \gamma_{j\to k}^{(1)}\gamma_{k\to f}^{(1)}+\beta_0\gamma_{j\to f}^{(1)}}{2}\log^2\frac{\mu^2}{k_\perp^2} + \left(
\beta_0 c_{j\to f}^{(1)} + \sum_k \gamma_{j\to k}^{(1)}c_{k\to f}^{(1)} + \gamma_{j\to f}^{(2)}
\right)\log\frac{\mu^2}{k_\perp^2}
\right]\nonumber\,,
\end{align}
where we have only included the necessary pieces to extract the two-loop anomalous dimension, $\gamma_{j\to f}^{(2)}$.  The renormalization factors through one-loop are
\begin{align}
Z_{q\to q}(\mu_0^2,\mu^2) &= 1+\frac{\alpha_sC_F}{2\pi}\left(
\frac{\mu_0^2}{\mu^2}
\right)^\epsilon\frac{1}{\epsilon}\left(
\frac{5}{8}-2\log 2\right)+\cdots\,,\\
Z_{q\to g}(\mu_0^2,\mu^2) &= \frac{\alpha_sC_F}{2\pi}\left(
\frac{\mu_0^2}{\mu^2}
\right)^\epsilon\frac{1}{\epsilon}\left(
-\frac{5}{8}+2\log 2\right)+\cdots\,,\\
Z_{g\to g}(\mu_0^2,\mu^2) &= 1-\frac{\alpha_s n_fT_R}{2\pi} \left(
\frac{\mu_0^2}{\mu^2}
\right)^\epsilon\frac{1}{\epsilon}
\frac{2}{3}+\cdots\,,\\
Z_{g\to q} (\mu_0^2,\mu^2)&= \frac{\alpha_sT_R}{2\pi} \left(
\frac{\mu_0^2}{\mu^2}
\right)^\epsilon\frac{1}{\epsilon}
\frac{1}{3}+\cdots\,.
\end{align}
The renormalized jet functions through one-loop are
\begin{align}
J^\text{(ren)}_{q\to q}(\mu^2,k_\perp^2) 
&=1+\frac{\alpha_sC_F}{2\pi} \left(
\frac{\mu_0^2}{k_\perp^2}
\right)^\epsilon\frac{1}{\epsilon}\left[
\frac{5}{8}-2\log 2+\epsilon\left(
2-\frac{7}{4}\log 2-2\log^2 2
\right)
\right]\\
&\hspace{1cm}-\frac{\alpha_sC_F}{2\pi}\left(
\frac{\mu_0^2}{\mu^2}
\right)^\epsilon\frac{1}{\epsilon}\left(
\frac{5}{8}-2\log 2\right)+{\cal O}(\alpha_s^2)\,,\nonumber\\
J^\text{(ren)}_{q\to g}(\mu^2,k_\perp^2) &=\frac{\alpha_sC_F}{2\pi} \left(
\frac{\mu_0^2}{k_\perp^2}
\right)^\epsilon\frac{1}{\epsilon}\left[
-\frac{5}{8}+2\log 2+\epsilon\left(
-2+\frac{7}{4}\log 2+2\log^22
\right)
\right]\\
&
\hspace{1cm}- \frac{\alpha_sC_F}{2\pi}\left(
\frac{\mu_0^2}{\mu^2}
\right)^\epsilon\frac{1}{\epsilon}\left(
-\frac{5}{8}+2\log 2\right)+{\cal O}(\alpha_s^2)\,,\nonumber\\ 
J^{(\text{ren})}_{g\to g}(\mu^2,k_\perp^2) &=1+\frac{\alpha_s n_fT_R}{2\pi} \left(
\frac{\mu_0^2}{k_\perp^2}
\right)^\epsilon\frac{1}{\epsilon}\left[
-\frac{2}{3}+\epsilon\left(
-\frac{17}{18}+\frac{4}{3}\log 2
\right)
\right] \\
&\hspace{1cm}
+\frac{\alpha_s n_fT_R}{2\pi}\left(
\frac{\mu_0^2}{\mu^2}
\right)^\epsilon\frac{1}{\epsilon}
\frac{2}{3}+{\cal O}(\alpha_s^2)\,,\nonumber\\
J^{(\text{ren})}_{g\to q}(\mu^2,k_\perp^2) &= \frac{\alpha_sT_R}{2\pi} \left(
\frac{\mu_0^2}{k_\perp^2}
\right)^\epsilon\frac{1}{\epsilon}\left[
\frac{1}{3}+\epsilon\left(
\frac{17}{36}-\frac{2}{3}\log 2
\right)
\right] -\frac{\alpha_sT_R}{2\pi} \left(
\frac{\mu_0^2}{\mu^2}
\right)^\epsilon\frac{1}{\epsilon}
\frac{1}{3}+{\cal O}(\alpha_s^2)\,.
\end{align}
The corresponding one-loop anomalous dimensions are
\begin{align}
\gamma_{q\to q} &= \frac{\alpha_s}{2\pi}C_F\left(
\frac{5}{8} - 2\log 2
\right)+{\cal O}(\alpha_s^2)\,,\\
\gamma_{q\to g} &= \frac{\alpha_s}{2\pi}C_F\left(
-\frac{5}{8} + 2\log 2
\right)+{\cal O}(\alpha_s^2)\,,\\
\gamma_{g\to q} &= \frac{\alpha_s}{2\pi}\frac{T_R}{3}+{\cal O}(\alpha_s^2)\,,\\
\gamma_{g\to g} &= -\frac{\alpha_s}{2\pi}\frac{2}{3}n_fT_R+{\cal O}(\alpha_s^2)\,.
\end{align}
The one-loop constants are
\begin{align}
c_{q\to q} &= \frac{\alpha_s}{2\pi}C_F\left(
2-\frac{7}{4}\log 2-2\log^22
\right)+{\cal O}(\alpha_s^2)\,,\\
c_{q\to g} &= \frac{\alpha_s}{2\pi}C_F\left(
-2+\frac{7}{4}\log 2+2\log^22
\right)+{\cal O}(\alpha_s^2)\,,\\
c_{g\to q} &= \frac{\alpha_s}{2\pi}T_R\left(
\frac{17}{36}-\frac{2}{3}\log 2
\right)+{\cal O}(\alpha_s^2)\,,\\
c_{g\to g} &= \frac{\alpha_s}{2\pi}n_fT_R\left(
-\frac{17}{18}+\frac{4}{3}\log 2
\right)+{\cal O}(\alpha_s^2)\,.
\end{align}
Finally, the coefficient of the $\beta$-function is
\begin{align}
\beta_0 = \frac{11}{6}C_A - \frac{2}{3}n_f T_R\,.
\end{align}

\bibliography{refs}

\end{document}